\documentclass[journal]{IEEEtran}

\usepackage{subfigure}
\usepackage{cite}

\ifCLASSINFOpdf
   \usepackage[pdftex]{graphicx}

\else

\fi

\usepackage[cmex10]{amsmath}
\usepackage{url}

\usepackage{amssymb}

\usepackage{epstopdf}

\begin{document}

\title{An Energy Efficient MAC Protocol for Fully Connected Wireless Ad Hoc Networks }

\author{Kamal~Rahimi Malekshan,~\IEEEmembership{Student member,~IEEE,}
        Weihua~Zhuang,~\IEEEmembership{Fellow,~IEEE}\\
        and~Yves~Lostanlen,~\IEEEmembership{Senior member,~IEEE}
\thanks{Manuscript received October 11, 2013; revised March 18, 2014; accepted June 29, 2014. The associate editor coordinating the review of this paper and approving it for publication was Y. Chen. This work was supported by a research grant from the Natural Sciences and Engineering Research Council (NSERC) of Canada. This paper was presented in part at IEEE ICC 2013 \cite{Kamal}.

K. Rahimi Malekshan and W. Zhuang are with the Department of Electrical and Computer Engineering,
University of Waterloo, Canada (e-mail: \{krahimim, wzhuang\}@uwaterloo.ca).

Y. Lostanlen is with SIRADEL North America, Toronto, Canada (e-mail: YLostanlen@siradel.com).

Digital Object Identiﬁer 10.1109/TWC.2014.2336801
}
}
\maketitle

\begin{abstract}
Energy efficiency is an important performance measure of wireless network protocols, especially for battery-powered mobile devices such as smartphones. This paper presents a new energy-efficient medium access control (MAC) scheme for fully connected wireless ad hoc networks. The proposed scheme reduces energy consumption by putting radio interfaces in the sleep state periodically and by reducing transmission collisions, which results in high throughput and low packet transmission delay. The proposed MAC scheme can also address the energy saving in realtime traffics which require very low packet transmission delay. An analytical model is established to evaluate the performance of the proposed MAC scheme. Analytical and simulation results demonstrate that the proposed scheme has a significantly lower power consumption, achieves substantially higher throughput, and has lower packet transmission delay in comparison with existing power saving MAC protocols.
\end{abstract}


\IEEEpeerreviewmaketitle

\section{Introduction}
\IEEEPARstart{T}{he} number of mobile devices and the demand for multimedia applications (which require high data rate and low packet transmission delay) have been increasing significantly in recent years. Wireless local area networks (WLANs), which have a shorter communication range in comparison with cellular networks, provide a more efficient and cost effective transmission for hot-spot mobile communications. In 2013, 45\% of total mobile data traffic was offloaded through WiFi or femtocell \cite{VNI}. A WLAN connects mobile devices in proximity and provides a link to the Internet through access points (APs). The set of mobile devices and APs form a wireless ad hoc network. Thus, effective mobile communication requires establishing energy efficient, high throughput, and low latency wireless ad hoc networks. The radio interface is a main source of energy consumption of mobile devices such as laptops and smartphones, which can quickly drain the device's limited battery capacity \cite{Tsao:2011:, Greening, Pering:2006:CRP:1134680.1134704,WiZi-Cloud}. For instance, the WiFi radio consumes more than 70\% of total energy in a smartphone when the screen is off \cite{Pering:2006:CRP:1134680.1134704}, which is reduced to 44.5\% and 50\% in the power saving mode when the screen is on and off respectively \cite{WiZi-Cloud}.

A radio interface can be in one of the following modes: transmit, receive, idle, and sleep. It has maximum power consumption in the transmit mode and minimum power consumption in the sleep mode. In the idle mode, a node needs to sense the medium and, hence, consumes a similar amount of power as in the receive mode. For instance, a \begin{it}Cisco Aironet 350 series\end{it} WLAN adapter \cite{Cisco} consumes 2.25W, 1.25W, 1.25W and .075W in the transmit, receive, idle and sleep modes respectively. Clearly, a significant amount of energy is consumed even in the idle mode. This occurs in the CSMA/CA\footnote{Carrier sense multiple access with collision avoidance.} mechanism in IEEE 802.11 \cite{Standard}, where each node in the network has to continually listen to the channel. To conserve energy, power saving mechanisms ‎\cite{Standard, SPAN, Rodoplu99minimumenergy, DPSM} allow a node to enter the sleep mode by powering off its radio interface when the node is not involved in transmission.

The medium access control (MAC) sub-layer determines how mobile nodes share the transmission medium in a wireless network. It also directly controls the operations of their radio interfaces. Therefore,  MAC plays an important role in achieving high throughput, low latency, and energy efficiency in wireless networks. Existing MAC protocols for wireless networks can be classified into contention-free and contention-based schemes. The former uses pre-defined assignments to allow nodes to transmit without contention, which includes time-division, polling, and token-based MAC protocols. In contention-based MAC, a node dynamically contends with other nodes to access the channel. Contention based schemes are more flexible and efficient in managing the channel in a distributed way when the data traffic load is not high. However, when the data traffic load is high, there are high chances of packet transmission collisions. The collisions cannot be detected quickly at the transmitting nodes, and the lack of an acknowledgment message is often the only way for the sender to detect collisions. Whenever a transmission collision happens, the radio bandwidth and the energy for transmitting and receiving a packet are wasted. Thus, an efficient MAC scheme should minimize the chances of transmission collisions to reduce energy and channel time wastage in a wireless network.

In this paper, we propose a novel MAC scheme for a fully connected wireless network. Using a temporary coordinator node, the proposed scheme reduces the energy consumption by scheduling the active and sleep times of node radio interfaces in a distributed way, and decreases MAC overhead and transmission collisions among nodes. A node contends only once to transmit a batch of packets, after that it will be assigned contention-free time for transmission by the temporary coordinator node, as long as it has packets ready for transmission. Nodes stay awake for a short time at the beginning of each beacon interval (to receive the transmission scheduling information) and during their packet transmission/reception times. The MAC scheme aims to satisfy delay and packet loss rate requirements and reduces energy consumption of nodes with realtime traffic such as voice or video calls that have stringent delay and packet loss requirements. Compared to existing power saving mechanisms, the proposed scheme has lower energy consumption, higher throughput, and shorter packet transmission delay.

The rest of this paper is organized as follows: Section II reviews related works. In Section III, we describe the proposed MAC protocol. Then, we present an analytical model to evaluate the performance of the proposed MAC scheme in Section IV. Numerical results are given in Section V to demonstrate performance of the proposed MAC scheme in comparison with existing MAC protocols. Finally, Section VI concludes this research.

\section{Related Works}
In this section, we review existing power saving MAC protocols proposed for wireless ad hoc networks and for networks with access point (AP) support.
\subsection{Power saving MAC for ad hoc networks}
The IEEE 802.11 DCF mode has a power saving mechanism (hereafter referred to as PSM) \cite{Standard}. In the ad hoc mode of PSM, time is partitioned into beacon intervals that are used to synchronize the nodes. All nodes must stay awake for a fixed time called Ad hoc Traffic Indication Message (ATIM) window at the beginning of each beacon interval. The ATIM window is used for nodes to announce the status of packets ready for transmission. The source nodes send ATIM frames to inform destinations that they have packets ready for transmission. The ATIM frames are acknowledged via ATIM-ACK packets during the same beacon interval transmitted from the destination nodes. Then, the sender and receiver nodes stay awake during the rest of the beacon interval for communication. Nodes use the CSMA/CA MAC protocol to access the channel during ATIM and the communication intervals.

In \cite{SPAN}, it is argued that only a fraction of the nodes (coordinators) need to be awake to forward the traffic for active connections. In the proposed algorithm, the other nodes (non-coordinators) follow the IEEE 802.11 PSM. There is a new window following the ATIM window, in which the non-coordinators (having announced packets in the ATIM window) exchange packets. After this newly introduced window, only coordinators stay awake to forward traffic. Zheng \begin{it}et al\end{it} \cite{On-demand} propose an on-demand power management scheme for multi-hop wireless ad hoc networks, which alternates between the PSM mode and active mode.

A fixed size ATIM window in PSM decreases performance in terms of throughput and energy consumption \cite{Woesner}. If the ATIM window size is too small, nodes do not have enough time to announce the buffered packets. On the other hand, when the ATIM size is too large, the packet transmission time is reduced and nodes cannot transmit the packets announced for transmission. In addition, a too large ATIM size increases the energy consumption, because all of the nodes have to stay awake during the ATIM interval. The dynamic power saving mechanism (DPSM) \cite{DPSM} dynamically adjusts the size of ATIM window based on the network condition to reduce energy consumption. In the DPSM, each node independently chooses the length of the ATIM window and dynamically changes it based on the network load condition.

In TMMAC \cite{TMMAC}, time is divided into beacon intervals and each beacon interval has an ATIM window which is used to announce the traffic similar to PSM. The CSMA/CA is used in ATIM window, but a contention-free MAC protocol is used for data packet transmission during the communication period. The communication period is divided into a fixed number of slots. The nodes reserve transmission slots by exchanging the control packets in the ATIM window.

An asynchronous power saving MAC is proposed for multi-hop ad hoc networks in \cite{quorum-based}. Each node has its own clock and in each beacon interval either stays awake for the whole beacon interval and transmits a beacon frame or follows the PSM mode of IEEE 802.11. The pattern of active and power saving sequence of beacon intervals are chosen based on a quorum algorithm that guarantees a beacon of a node will be heard by any other node at least twice in a given interval of $N$ beacons. The clock and wake up pattern of the sender node are included in its beacon frame that allows the other nodes to become aware of the nodes wake up time to successfully deliver packets.

In \cite{RTS-CTS}, a power saving mechanism is proposed based on the IEEE 802.11 RTS/CTS\footnote{Request-to-Send/Clear-to-Send} dialogue. Nodes that are neither transmitting nor receiving a packet switch off their wireless interfaces after overhearing RTS/CTS packets. In \cite{PAMAS}, all the RTS/CTS packet exchanges are performed over the signaling channel and data packets are transmitted over a different channel, which allows nodes to independently decide whether to stay awake or power off after overhearing the RTS/CTS packets.

\begin{figure*}[t]
  \centering
  \includegraphics[width=7.4in, trim=0.5cm 12cm 0cm 6.4cm, clip=true]{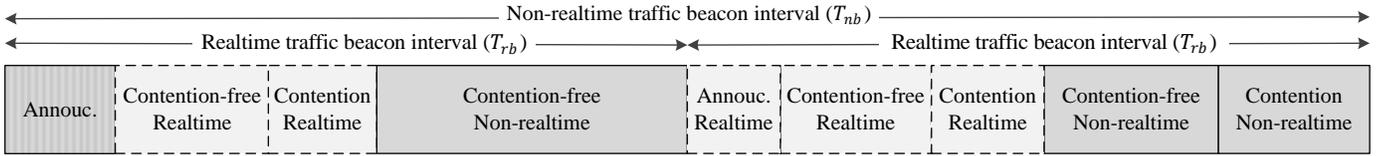}\\
  \caption{Structure of one beacon interval of the proposed scheme for $\beta=2$.}
  \label{FrameN2}
\end{figure*}

\subsection{Power saving MAC for networks with APs}
The PSM in a network with AP support is similar to that in an ad hoc network. Time is portioned to beacon intervals. At beginning of each beacon interval, the AP broadcasts a Traffic Indication Message (TIM) to inform a power saving node of packets ready for it in the AP. The nodes that have packets ready for them in the AP stay awake during the rest of beacon interval and poll the AP to receive the packets. Nodes that have packets ready for transmission to the AP also stay awake and send their packets to the AP during the beacon interval \cite{Standard}. The PSM performance is improved in energy saving by separating the delay sensitive traffic and delay tolerant traffic \cite{catnap}, giving priority to the power saving nodes \cite{NAPman}, and distributing TIM of different APs to avoid traffic burst \cite{SleepWell}. The PSM performance for delay sensitive applications is investigated and active/sleep schedules that guarantees delay requirements are proposed in \cite{Self-tuning,SPSM,VoIP-PowerSaving,Bounded-slowdown}. Other power saving protocols have been proposed in literature which require another low power interface (e.g. ZigBee) along WiFi interface (e.g. \cite{Pering:2006:CRP:1134680.1134704} and \cite{WiZi-Cloud}) or require physical layer modifications (e.g. \cite{Micro-Power-Management} and \cite{E-MiLi}).

The power saving MAC protocols for networks with AP support can provide high performance and low energy consumption when there is only one AP in the network. However, in wireless local area networks, several APs are usually located in the same area and have to contend with each other to access the shared channels, which degrades the network throughput and increases the energy consumption. In fact, the set of APs and nodes connected to APs form a wireless ad hoc network.

In this work, we propose a new power saving scheme that has significantly lower power consumption than existing protocols for a fully connected wireless ad hoc network. The proposed scheme adapts to the network conditions and it has smaller contention overhead and power saving mechanism overhead. Thus, it provides substantially higher throughput and lower packet transmission delay. In addition, the proposed protocol aims to satisfy the delay and packet loss rate requirements and reduce energy consumption of nodes with realtime traffic.

\section{Proposed MAC Protocol}
Consider a single-channel fully connected wireless network, consisting of $N$ nodes with realtime traffic and $K$ nodes with non-realtime traffic, where all the nodes can hear transmissions from each other. There is no dedicated central controller in the network.

Time is partitioned into beacon intervals of constant duration and all nodes are synchronized in time. The synchronization can be achieved by using a distributed beacon transmission mechanism, as in the IEEE 802.11 power saving mechanism \cite{Standard}.

Each beacon interval consists of three different periods: \begin{it}announcement period\end{it}, \begin{it}contention-free period\end{it}, and \begin{it}contention period\end{it}. Figure \ref{FrameN2} shows the structure of one beacon interval. The durations of the periods are adjusted dynamically by a temporary coordinator node called \begin{it}head node\end{it}, based on the instantaneous network traffic load condition. The \begin{it}head node\end{it} monitors the traffic demands of nodes in the previous beacon interval and records nodes' requests in a table called \begin{it}demand table\end{it}. At the beginning of the current beacon interval, in the \begin{it}announcement period\end{it}, the \begin{it}head node\end{it} broadcasts the durations of the periods, and transmission schedule of the \begin{it}contention-free period\end{it} based on the \begin{it}demand table\end{it}. The current \begin{it}head node\end{it} also randomly selects one of the rest nodes as the new \begin{it}head node\end{it} for the next beacon interval. All nodes must be awake to receive the broadcast message in the \begin{it}announcement period\end{it} from the current \begin{it}head node\end{it}. The scheduled transmissions take place in the \begin{it}contention-free period\end{it} with SIFS\footnote{The Short Inter-frame Space (SIFS) equals to time required for a node to sense the end of a packet transmission and start transmitting.} intervals in between. Nodes, which have packets to transmit but have not informed the \begin{it}head node\end{it} of their intent for transmission, contend in the \begin{it}contention period\end{it} using a CSMA MAC mechanism to inform the \begin{it}head node\end{it} of their intention for transmission by sending an RTS packet.

In the proposed MAC protocol, nodes need to wait for one beacon interval to transmit packets. Since realtime traffic (from voice and video calls) has a strict packet transmission delay requirement, the beacon interval for realtime traffic should be less than the maximum tolerable packet delay of realtime traffic. On the other hand, non-realtime traffic (such as in file transfer and web browsing) can tolerate a longer packet transmission delay, and a small beacon interval increases the energy consumption because nodes have less sleep times. Therefore, we propose different beacon intervals for realtime traffic and non-realtime traffic respectively. We also assign unique transmission time to the realtime traffic for higher priority over non-realtime traffic flows.

Let $D_{M}$ denote the maximum tolerable packet delay of realtime traffic. The beacon interval duration of realtime traffic, $T_{rb}$, should not be longer than $D_M$, $T_{rb}\leq D_M$. We set the beacon interval duration for non-realtime traffic as, $T_{nb}=\beta T_{rb}$, where $\beta \geq 1$ is an integer. Figure \ref{FrameN2} shows the structure of the realtime and non-realtime beacon intervals for $\beta=2$. That is, there are two realtime beacon intervals per non-realtime beacon interval. As $\beta$ is increased, the throughput is increased and the energy consumption in nodes with non-realtime traffic is reduced due to less frequent \begin{it}announcement period\end{it}s, however, the packet transmission delay is increased because of larger non-realtime beacon intervals. The realtime traffic frame duration in a realtime beacon interval ($T_{rf}$), which is the summation of \begin{it}contention-free period\end{it} and \begin{it}contention period\end{it} of realtime traffic in one realtime beacon interval, is constant and should be adjusted to meet the packet loss rate requirement of realtime traffic in the network. The value of parameters $T_{rb}$, $T_{rf}$, and $\beta$, can be updated by the \begin{it}head node\end{it} based on the network condition. Table \ref{Table0} summarizes the notations. In the following, we discuss detail operation of the proposed MAC protocol in a non-realtime beacon interval.

\begin{it}Announcement period\end{it}s: There are $\beta$ \begin{it}announcement period\end{it}s per one non-realtime beacon interval. In each period, the \begin{it}head node\end{it} regulates the transmission for the current beacon interval and announces the transmission schedule by broadcasting a \begin{it}scheduling packet\end{it}, based on the requests in the \begin{it}demand table\end{it} that it generated/updated in the previous beacon interval. When the number of requests is more than the packet transmissions that can be scheduled in the current beacon interval, the \begin{it}head node\end{it} also broadcasts the pending requests. In the first \begin{it}announcement period\end{it} at the beginning of each non-realtime beacon interval, all the nodes (with realtime and non-realtime traffic) stay awake and the \begin{it}head node\end{it} schedules the transmission for the current non-realtime beacon interval and the first realtime beacon interval. The \begin{it}head node\end{it} also randomly selects one of the nodes that is involved in transmission/reception of current beacon interval as the new \begin{it}head node\end{it} for the next beacon interval. The \begin{it}scheduling packet\end{it} in the first \begin{it}announcement period\end{it} contains the following information: the duration of realtime frames and the starting times of next $\beta-1$ \begin{it}announcement periods\end{it}, scheduling information for the first realtime beacon interval, the scheduling information for the non-realtime traffic in the current non-realtime beacon interval, and the node selected as the next \begin{it}head node\end{it}. The scheduling information for the first realtime beacon interval determines the transmission/reception in the first realtime \begin{it}contention-free period\end{it}, the duration of realtime \begin{it}contention period\end{it}, and the pending requests that cannot be scheduled in the first realtime beacon interval. The scheduling information for the non-realtime traffic determines the transmission/reception schedule in the non-realtime \begin{it}contention-free period\end{it}, the duration of non-realtime \begin{it}contention period\end{it}, and the pending requests that cannot be scheduled in the current non-realtime beacon interval\footnote{Note that the \begin{it}contention-free period\end{it} and the \begin{it}contention period\end{it} of non-realtime traffic may have more than one parts which are separated by realtime frames.}. The node selected as the next \begin{it}head node\end{it} should confirm with an ACK packet following the \begin{it}scheduling packet\end{it}. If the selected node does not confirm, the \begin{it}head node\end{it} will continue to serve as the \begin{it}head node\end{it} for the next non-realtime beacon interval, and then will select a different \begin{it}head node\end{it} at the first \begin{it}announcement period\end{it} of the next non-realtime beacon interval. In the next ($\beta-1$) \begin{it}announcement period\end{it}(s) of the current non-realtime beacon interval, only the nodes with realtime traffic are awake and the \begin{it}head node\end{it} schedules the transmission/reception in the realtime beacon intervals based on the transmission requests in the previous realtime beacon interval. The \begin{it}scheduling packet\end{it}, transmitted by the \begin{it}head node\end{it} in each \begin{it}announcement period\end{it} at the beginning of next $\beta-1$ realtime beacon intervals, contains the transmission/reception schedule for that \begin{it}realtime contention-free period\end{it}, and the pending requests that cannot be scheduled in that realtime beacon interval.

\begin{it}Contention-free period\end{it}s: In these periods, the \begin{it}head node\end{it} stays awake and the transmitter/receiver nodes that are scheduled for transmitting/receiving packets wake up at the assigned time to transmit/receive packets. The nodes with realtime traffic are scheduled to transmit/receive packets at the \begin{it}realtime contention-free period\end{it}s and the nodes with non-realtime traffic are scheduled to transmit/receive the data packets in the \begin{it}non-realtime contention-free period\end{it}s. Sender nodes with realtime traffic put their call status (\textit{on} or \textit{off}) in the header of the transmitted packets, and the sender nodes with non-realtime traffic put the number of the remaining packets ready for transmission in the header of their data packets. The \begin{it}head node\end{it} uses the information to generate/update the \begin{it}demand table\end{it}. Although transmission of packets is collision free in the \begin{it}contention-free period\end{it}, the transmission may be corrupted by short-term channel fading. Therefore, receivers should acknowledge receiving non-realtime packets by transmitting an ACK packet\footnote{Note that instead of transmitting an individual ACK for each packet, multiple packets can be acknowledged using a single Block ACK \cite{Standard} to improve the MAC efficiency.}. In contrast, realtime packets will be useless if they are not transmitted before a deadline. Thus, no ACK packet is transmitted by the receiver for realtime packets.

\begin{it}Contention period\end{it}s: In the \begin{it}contention period\end{it}s, nodes (that have packets ready for transmission but were neither scheduled for transmission nor included in the pending traffic list transmitted by the \begin{it}head node\end{it} in the previous \begin{it}{announcement period}\end{it}) stay awake and contend for transmission using a CSMA MAC protocol to submit a transmission request. Nodes with realtime traffic submit transmission requests at the \begin{it}realtime contention period\end{it}s and node with non-realtime traffic submit transmission requests at the \begin{it}non-realtime contention period\end{it}s. Once a contending node's back-off counter reaches zero, it transmits an RTS packet to the \begin{it}head node\end{it}. The RTS packet of a node with realtime traffic includes information of the maximum tolerable delay, maximum tolerable packet loss rate, the sender node ID, and the destination node ID. The RTS packet of a node with non-realtime traffic contains the number of packets that are ready for transmission at the sender, the sender node ID, and the receiver node ID. The \begin{it}head node\end{it} stays awake to monitor transmission requests and records them in the \begin{it}demand table\end{it}. When a node successfully transmits a request without collision, the \begin{it}head node\end{it} records the information in the \begin{it}demand table\end{it} and uses this information to schedule transmission at the next beacon interval. Once a contending node has submitted a request to the \begin{it}head node\end{it}, it powers off for the rest of the beacon interval. If a contending node does not have a chance to submit a request, it will contend again in the \begin{it}contention period\end{it} of the next beacon interval.

\begin{table}[tp]
    \begin{center}
    \caption{Summary of Notations}
     \label{Table0}
        \begin{tabular}{cl}

          \hline
          \hline
          \begin{bf}Symbol\end{bf} & \begin{bf}Explanation\end{bf}\\
          \hline

            $D_M$ &        Maximum tolerable delay of a realtime packet\\
            $h$ &          Payload of a realtime packet\\
            $K$ &          Number of the nodes with non-realtime traffic\\
             $M$&          Maximum number of sender nodes with realtime  \\
                &          traffic that can be scheduled for transmission  \\
              &            in one realtime beacon interval\\
            $m(s)$ &       Number of nodes with realtime traffic scheduled \\
              &            for transmission when the system state is $s$\\
            $N$ &          Number of the nodes with realtime traffic\\
            $N_i$ &        Number of sender nodes with realtime traffic  \\
              &            that are in state $i\in\{1,2,...,5\}$ at the beginning \\
              &            of each beacon interval \\
            $S=$&          System state at each beacon interval \\
            $(N_1,N_2,N_3,N_4)$&            \\

            $T_{cf}(s)$ &  Duration of contention-free period when \\
             &             the system state is $s$\\
            $T_{cp}(s)$ &  Duration of contention period when the system \\
             &             state is $s$\\
            $T_{rb}$ &     Beacon interval for realtime traffic\\
            $T_{nb}$ &     Beacon interval for non-realtime traffic\\
            $T_{rf}$ &     Realtime traffic frame duration in a realtime \\
                     &     beacon interval\\
            $t_{on}$ &     Average duration of \textit{on} period of a realtime call\\
            $t_{off}$&     Average duration of \textit{off} period of a realtime call\\
            $t_{a}$ &      Inter-arrival time of packets in the \textit{on} mode   \\
                    &      of a realtime call\\

            $X_i$ &        Number of nodes that have transition  \\
             &             $i\in\{1,2,...,8\}$ in a realtime beacon interval\\
            $\delta^*$&    Maximum tolerable packet loss rate of each \\
                    &      realtime call\\
            $\delta_{ch}$& Packet loss rate due to channel impairments\\
            $\delta_{mac}$&Packet loss rate due to MAC contention\\
            $\rho$ &       Payload of an aggregated realtime packet\\
            $\tau_q$ &     Duration of one transmission request packet\\
            $\tau_v$ &     Duration of one aggregated realtime packet \\
          \hline
           \hline
        \end{tabular}
    \end{center}

\end{table}
The proposed scheme dynamically adjusts the transmission schedule of the periods based on the current traffic load condition of all nodes. It has the following features:

\subsubsection{The awake time of the nodes is short which reduces energy consumption}
Nodes with non-realtime traffic that are not involved in transmission/reception stay awake only at the first \begin{it}announcement period\end{it} at each non-realtime beacon interval. Also, nodes with realtime traffic stay awake only at the \begin{it}announcement period\end{it}s in each non-realtime beacon interval. The nodes that are scheduled to transmit or receive a packet wake up at the assigned time to transmit/receive without contention in the \begin{it}contention-free period\end{it}s. In the \begin{it}contention period\end{it}s, only the nodes that want to initiate a new transmission and the \begin{it}head node\end{it} stay awake. The \begin{it}head node\end{it} is the only node that stays awake for the whole beacon interval;

\subsubsection{The contention and collision overhead is small, which reduces the energy consumption and enhances the network performance}
Nodes contend for the channel only when they want to initiate a new transmission. Once a node successfully submits a transmission request, it will be assigned a transmission time in the next beacon intervals as long as it has packets ready for transmission. Also, the number of contending nodes decreases because each node does not content for transmission of each packet, but for transmission of a batch of packets available in its buffer.

\subsubsection{It is distributed and the scheduling workload is shared among all nodes in the network}
No dedicated central controller is required to manage the network. Nodes cooperate and in each beacon interval a coordinator node (\begin{it}head node\end{it}) schedules the transmissions based on the transmission requests from all nodes in the previous beacon interval. Nodes take turn to become the head node, such that the head of current beacon interval randomly selects one of the rest nodes as the new head node for the next beacon interval. Thus, the energy consumption at the head node for scheduling transmissions (i.e., energy consumed at the head node to stay awake for monitoring transmission requests in a beacon interval and transmit a scheduling packet) is distributed among all nodes in the network.

\subsubsection{It efficiently utilizes radio channel time and provides fair channel access}
In the proposed MAC, the \begin{it}head node\end{it} monitors transmission request of nodes for a beacon interval before scheduling transmissions/receptions at the beginning of subsequent beacon interval. The dedicated duration for \begin{it}contention period\end{it} is greater than a minimum value in every beacon interval, which ensures that every node can submit its transmission request to the head node even if the network is overloaded. Therefore, the head node has the information of all requests and can efficiently schedule the transmissions based on the instantaneous network load condition. Fair channel access can also be achieved, for instance, by using a Round Robin scheduling algorithm \cite{Round-Robin} at the head node.

In the following section, we present an analytical model to evaluate the performance of the proposed MAC protocol. The analytical model enables us to determine the minimum required frame time for realtime traffic in each realtime beacon interval to meet the packet loss rate requirements of realtime traffic flows.

\section{Performance Analysis for Realtime Traffic}
Time is discretized and normalized to the duration of a mini-slot\footnote{A mini-slot is the summation of RxTx turn around time, channel sensing time, propagation delay, and MAC processing delay.}. Let $N$ denote the number of nodes with realtime traffic in the network. The destination for each source node is selected randomly from the rest nodes. Consider constant rate, on-off realtime (voice or video) calls to make the analysis tractable. The duration of \textit{on} and \textit{off} modes are exponentially distributed with average $t_{on}$ and $t_{off}$ respectively. Packets are generated periodically with inter-arrival time $t_a$ in the \textit{on} mode, while no packet is generated in the \textit{off} mode. Each packet has a payload of $h$ bits. A sender node with realtime traffic aggregates the packets and transmits them as one packet at the assigned time in the \begin{it}realtime contention-free period\end{it}s. The payload of an aggregated packet is $\rho=\frac{T_{rb}}{t_a}h$. If an aggregated realtime packet is not transmitted within a deadline $D_M$, it will be removed at the sender. The maximum tolerable packet loss rate for each realtime call is $\delta^*$. Let $\delta_{ch}$ denote the packet error rate due to channel impairments and $\delta_{mac}$ denote the packet loss rate due to MAC contentions. The packet loss rate of each realtime call is given by
\begin{equation}\label{27}
  \delta=1-(1-\delta_{mac})(1-\delta_{ch}).
\end{equation}

According to (\ref{27}), the maximum allowable packet loss rate due to MAC contentions, $\delta_{mac}^*$, is
\begin{equation}\label{28}
  \delta_{mac}^*=1-\frac{1-\delta^*}{1-\delta_{ch}}.
\end{equation}

Let $T_{rf}$ denote the realtime traffic frame duration which is the summation of \begin{it}contention-free period\end{it} and \begin{it}contention period\end{it} assigned to realtime traffic in each realtime beacon interval. Let $\tau_q$ denote the duration of one transmission request packet (including an SIFS) and $t_v$ denote the transmission time of one aggregated realtime packet (including an SIFS) over the channel. The maximum number of nodes with realtime traffic that can be scheduled for transmission in one realtime beacon interval is
\begin{equation}\label{7}
  M=\lfloor\frac{T_{rf}}{t_v}\rfloor
\end{equation}
where $\lfloor .\rfloor$ denotes the floor function.

\begin{figure}[t]
\centering
\includegraphics[width=3.in, trim=4cm 4.5cm 6cm 6cm, clip=true]{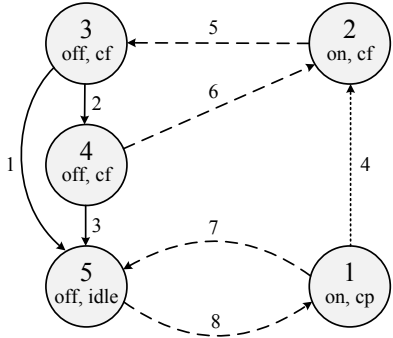}
\caption{Sender node states of a realtime call.}
\label{Status}
\end{figure}

\subsection{Markov modeling of the system}
At any beacon interval, the sender node of a realtime call is in one of the following states:\\

\begin{enumerate}
  \item State 1 -- The realtime call is in the \textit{on} mode but the sender node is not included in the \begin{it}demand table\end{it}. Thus, the sender node contends with other nodes in the \begin{it}realtime contention period\end{it} to submit a transmission request;

  \item State 2 -- The realtime call is in the \textit{on} mode and the sender node is included in the \begin{it}demand table\end{it};

  \item State 3 -- The realtime call has just switched to the \textit{off} mode, the sender node is included in the \begin{it}demand table\end{it} and has a pending packet for transmission whose transmission delay threshold has not passed yet. The sender node will inform its status change to the \begin{it}head node\end{it} when transmitting in the \begin{it}contention-free period\end{it};

  \item State 4 -- The realtime call is in the \textit{off} mode, and the sender node is included in the \begin{it}demand table\end{it}. However, the sender node has no pending packet for transmission. The sender node will inform the \begin{it}head node\end{it} of its \textit{off} mode when transmitting in the \begin{it}contention-free period\end{it};

  \item State 5 -- The realtime call is in the \textit{off} mode, and the node is not included in the \begin{it}demand table\end{it}.
\end{enumerate}

Figure \ref{Status} shows the sender node states of a realtime call. The state transitions illustrated by solid lines (transitions $1$, $2$ and $3$) take place in the \begin{it}contention-free period\end{it}. The transition represented by dotted line (transition $4$) takes place in the \begin{it}contention period\end{it}. The set of transitions indicated by dashed lines (transitions $5$, $6$, $7$ and $8$) are due to status changes of realtime call (from \textit{on} to \textit{off}, or from \textit{off} to \textit{on} mode) that we assume take place at the end of each realtime beacon interval. We assume that status of a realtime call does not change more than once during a realtime beacon interval, which is reasonable as average \textit{on} and \textit{off} periods of a realtime traffic call are in general much larger than the realtime beacon interval.

Let $N_i$ denote the number of sender nodes with realtime traffic that are in state $i\in\{1,...,5\}$ at the beginning of each beacon interval. We have $N_5=N-\sum_{i=1}^{4}N_i$. Denote the system state at each realtime beacon interval by $S=\big(N_1,N_2,N_3,N_4\big)$. Let $\mathcal{S}$ be the set of feasible system states in any beacon interval, $\mathcal{S}:\{s=(n_1,n_2,n_3,n_4)| n_i\geq 0,\sum_{i=1}^4 n_i\leq N\}$. When the system is in state $s=\big(n_1,n_2,n_3,n_4\big)$, the number of nodes with realtime traffic that are scheduled for transmission at that beacon interval is
\begin{equation}\label{90}
  m(s)=\min(n_2+n_3+n_4,M)
\end{equation}
and the corresponding durations of \begin{it}contention-free period\end{it} $T_{cf}(s)$ and \begin{it}contention period\end{it} $T_{cp}(s)$ are
\begin{equation}\label{91}
  T_{cf}(s)=m(s)\tau_v, \text{       } T_{cp}(s)=T_{rf}-T_{cf}(s).
\end{equation}

For a given number of realtime calls (no new call arrival and no call departures), since the duration of \textit{on} and \text{off} periods are exponentially distributed, given the current system state, all state transitions during the current beacon interval are independent of the system states in the previous beacon intervals. As the system state in the next beacon interval only depends on the system state in the current beacon interval and the number of transitions during the current beacon interval, the system state sequence satisfies the Markov property and is stationary. In the following subsection, we calculate the steady state probability of system states.

\subsection{Steady state probability of system states}
Let random vector $X=(X_1,...,X_8)$ denote the number of transitions during a realtime beacon interval, where $X_i$, $i\in\{1,...,8\}$, is the number of nodes that have state transition $i$ during the beacon interval. Let $\mathcal{X}(s,s')$ be the set of number of transitions $(x_1,...,x_8)$ during a beacon interval that change the system state from $s=(n_1,n_2,n_3,n_4)$ to $s'=(n_1',n_2',n_3',n_4')$. We have
\begin{multline}\label{23a}
\mathcal{X}(s,s')= \left\{(x_1,...,x_8)\left|
 \begin{array}{l}
  x_8-x_7-x_4=n_1'-n_1;\\
  x_4-x_5+x_6=n_2'-n_2;\\
  x_5=n_3';\\
  x_1+x_2=n_3;\\
  x_2-x_3-x_6=n_4'-n_4.
 \end{array}
\right
.\right\}.
\end{multline}
The probability of transition from system state $s$ to system state $s'$ in a beacon interval is
\begin{equation}\label{23}
  P_{s,s'}=\sum_{\mathcal{X}(s,s')} P_{X_1...X_8|S}(x_1,...,x_8|s)
\end{equation}
where $P_{X_1...X_8|S}(x_1,...,x_8|s)$ is the probability mass function (pmf) of the state transition numbers during a beacon interval given the system state $s$. Using conditional probability,
\begin{multline}\label{2b}
  P_{X_1...X_8|S}(x_1,...,x_8|s)=P_{X_1X_2X_3|S}(x_1,x_2,x_3|s)\\
  P_{X_4|X_1X_2X_3S}(x_4|x_1,x_2,x_3,s)\\ P_{X_5...X_8|X_1...X_4S}(x_5,...,x_8|x_1,...,x_4,s).
\end{multline}
In the right side of (\ref{2b}), the first term denotes the conditional pmf of state transition number (transitions $1$, $2$, $3$) at the \begin{it}contention-free period\end{it} given system state $s$. To calculate this term, we need to find the pmf of the node numbers in states $3$ and $4$ that are scheduled for transmission in the \begin{it}contention-free period\end{it}. The second term in the right side of (\ref{2b}) is the conditional pmf of the state transition number (transition $4$) at the \begin{it}contention period\end{it} given system state $s$. This can be obtained by analysing the CSMA MAC protocol to find the pmf of the number of successful transmission requests in the \begin{it}contention period\end{it}. The last term in the right side of (\ref{2b}) denotes the conditional pmf of the state transition numbers (transitions $5$, $6$, $7$, and $8$) due to status change of realtime calls given system state $s$, which can be found based on the distribution of $on$ and $off$ modes of realtime calls. We derive analytical expressions for these terms in Appendices A-C.

Finally, the steady state probability of the system states, $\pi(s), s\in\mathcal{S}$, can be found based on the transition probability between states given in (\ref{23}) using the balance equations.

\subsection{Minimum frame duration to guarantee the required QoS of realtime traffic}
When the number of nodes in the \begin{it}contention-free period\end{it} is more than $M$ and/or when nodes do not get a chance to successfully submit a transmission request in the \begin{it}contention period\end{it}, packet loss occurs. Although nodes in state $4$ may be scheduled for transmission, they do not have a packet for transmission. Therefore, when the system is in state $s=(n_1,n_2,n_3,n_4)$, the average number of transmitted packets in one realtime beacon interval is
\begin{equation}\label{25}
\bar{r}(s)=\min(n_2+n_3+n_4,M)\frac{n_2+n_3}{n_2+n_3+n_4}\rho.
\end{equation}
Nodes in states $1$ and $2$ are in the \textit{on} mode. Thus, the number of packets generated in one realtime beacon interval in the system state $s=(n_1,n_2,n_3,n_4)$ is
\begin{equation}\label{26a}
  \bar{g}(s)=(n_1+n_2)\rho.
\end{equation}
Using (\ref{25}) and (\ref{26a}), the packet loss rate due to MAC contention can be calculated as
\begin{equation}\label{26}
  \delta_{mac}=1-\sum_{s\in \mathcal{S}}\pi(s)\frac{\bar{r}(s)}{\bar{g}(s)}.
\end{equation}
To meet the required packet loss rate, the minimum frame time for realtime traffic $T_{rf}^*$ can be calculated by solving the following optimization problem,
\begin{equation}\label{29}
  \begin{array}{ll}
  T_{rf}^*&=\min T_{rf} \\
  &\text{s.t.   }\delta_{mac} \leq \delta_{mac}^*.
  \end{array}
\end{equation}
Since the packet loss rate due to MAC contention ($\delta_{mac}$) is a decreasing function of the dedicated time ($T_{rf}$) to realtime traffic in each realtime beacon interval, the optimization problem (\ref{29}) can be solved using the binary search algorithm.

\section{Numerical Results and Discussions}
Similar to the IEEE standard \cite{Standard}, realtime and non-realtime packets are transmitted with the data channel rate and all control packets (including ATIM, ATIM-ACK, RTS, ACK, and the \begin{it}scheduling packet\end{it}) are transmitted using the basic channel rate. The destination node for each source node is selected randomly from the rest nodes. We use 2.25W, 1.25W, 1.25W and .075W as values of power consumption by each radio interface in the transmit, receive, idle, and sleep states respectively, based on the data of \begin{it}Cisco Aironet Wireless LAN Adapters 350 series\end{it} \cite{Cisco}. Simulations are performed using MATLAB for 100 seconds of the channel time, with error-free transmissions.

\begin{table}[tp]
    \begin{center}
    \caption{Simulation Parameters}
     \label{Table}
        \begin{tabular}{l|l}
          \hline
          \hline
          Parameter & Value  \\
           \hline
          Slot time &     $20$ $\mu$s  \\
          SIFS      &      $10$ $\mu$s   \\
          DIFS      &      $50$ $\mu$s   \\
          $W$ &    32   \\
          $CW_{min}$ &    $15$    \\
          $CW_{max}$ &    $1023$  \\
          PHY preamble & $192$ $\mu$s   \\

          RTS size & $160$ bits \\
          CTS size & $112$ bits \\
          ACK size & $112$ bits \\
          ATIM size & $224$ bits] \\
          ATIM-ACK size & $112$ bits \\
          Scheduling size for one transmission &  $160$ bits   \\
          Non-Realtime Beacon interval  & $100$ $ms$ \\
          Realtime Beacon interval  & $50$ $ms$ \\
          Data rate     &  $11$ Mbps  \\
          Basic rate     &      $2$ Mbps \\
          \hline
          $t_{on}$ & $1.8$ seconds\\
          $t_{off}$ & $1.2$ seconds\\
          Voice codec & G.711 (64Kbps)  \\
          Voice packet inter arrival time& $20$ $ms$\\
          Voice packet payload & $160$ bytes\\
          User datagram protocol (UDP) overhead & $8$ bytes\\
          Realtime transport protocol (RTP) overhead & $12$ bytes\\
          IP overhead & $20$ bytes\\
          MAC overhead & $20$ bytes\\
          Maximum voice packet delay & $50$ $ms$\\
          Maximum packet loss rate of voice traffic & $1$\%\\
          Data packet size & $1024$ bytes  \\

          \hline
           \hline
        \end{tabular}
    \end{center}

\end{table}

\begin{figure*}
\centering
\subfigure[K=10]{\includegraphics[width=2.31in, trim=.4cm 1.4cm 1cm 1cm, clip=true]{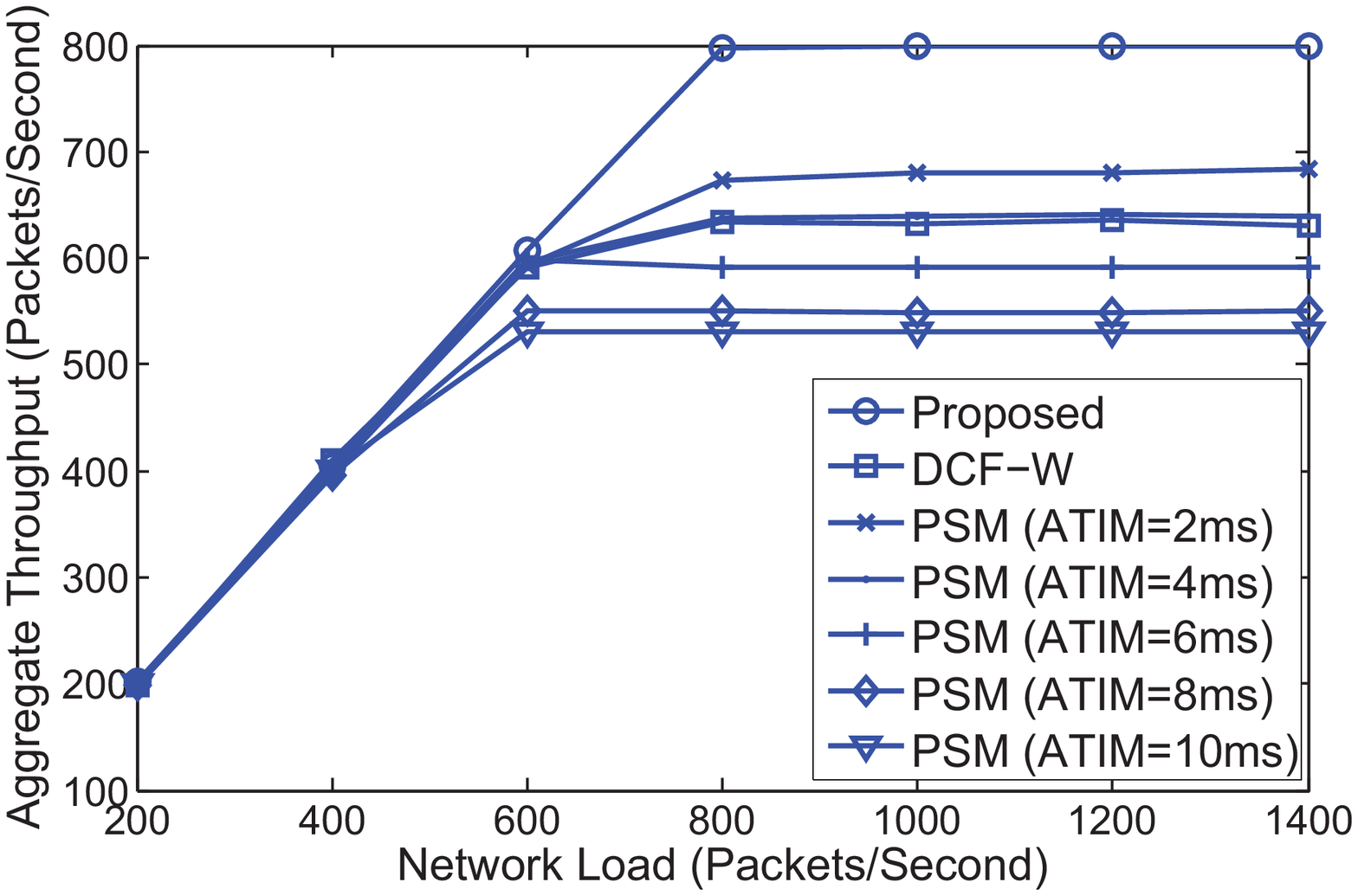}\label{T1}}
\subfigure[K=20]{\includegraphics[width=2.31in, trim=.4cm 1.4cm 1cm 1cm, clip=true]{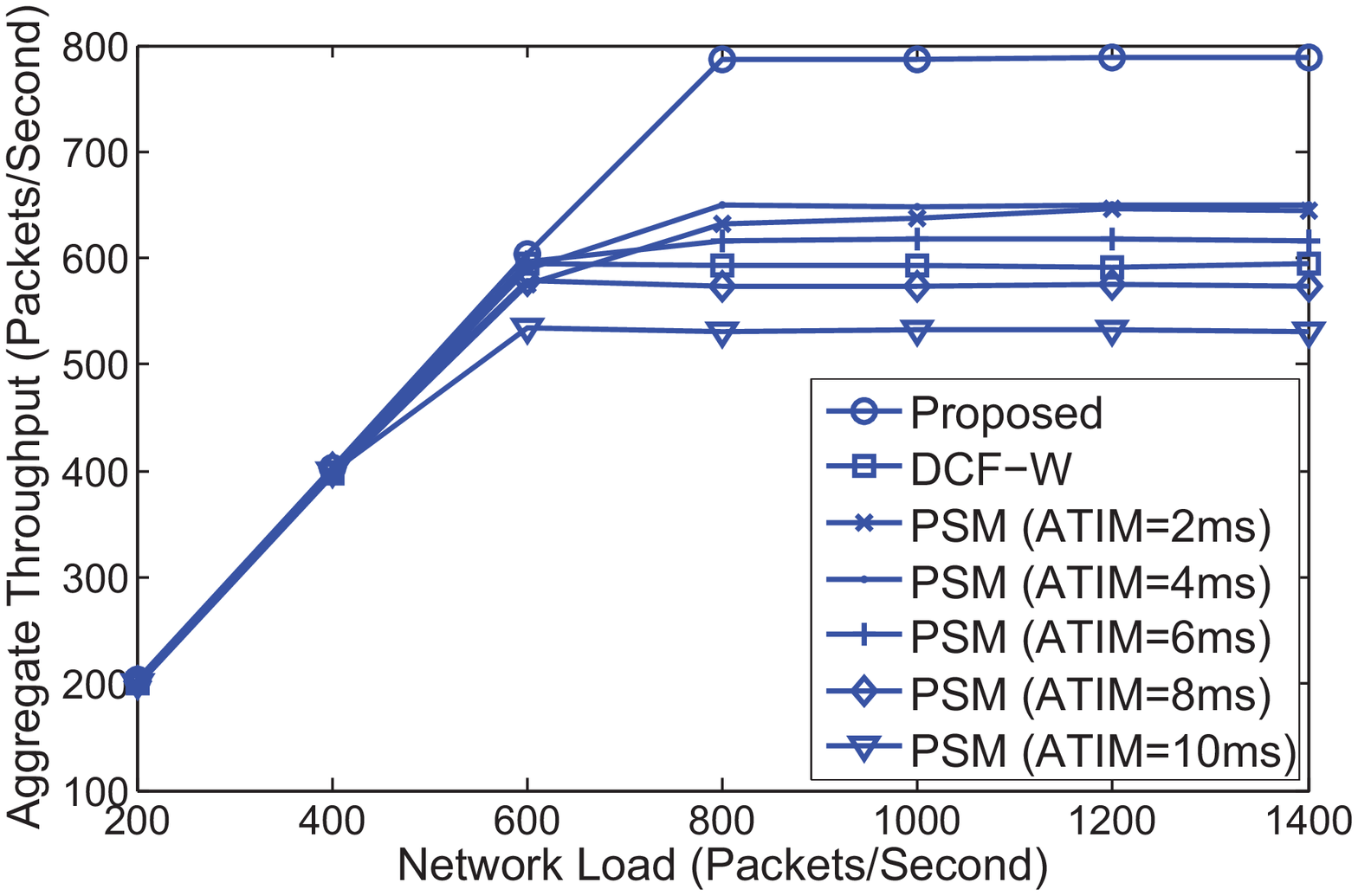}\label{T2} }
\subfigure[K=50]{\includegraphics[width=2.31in, trim=.4cm 1.4cm 1cm 1cm, clip=true]{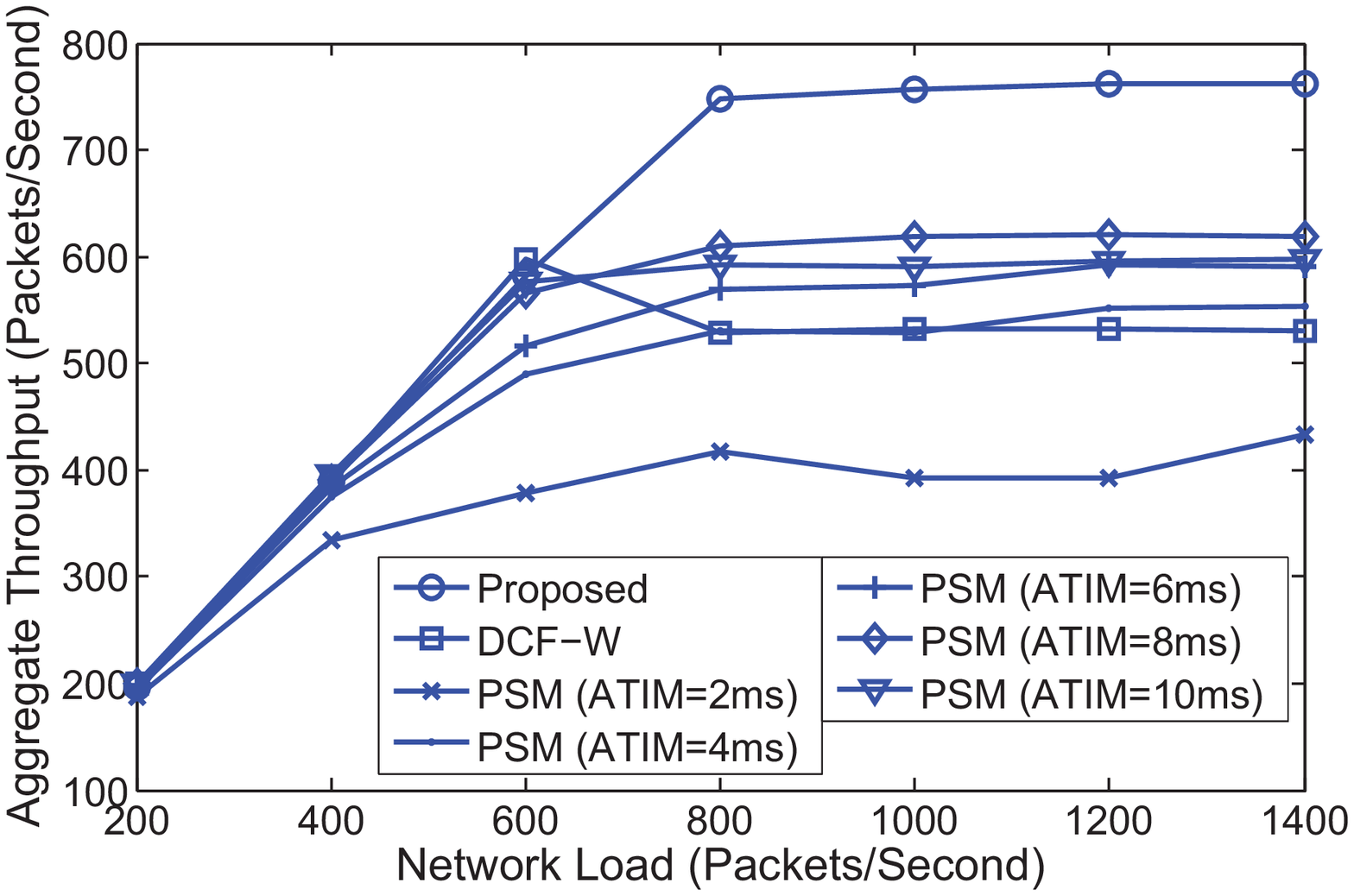} \label{T5}}
\caption{Aggregate throughput of the proposed scheme, PSM, and DCF-W.} \label{T}
\end{figure*}
\begin{figure*}
\centering
\subfigure[K=10]{\includegraphics[width=2.31in, trim=.4cm 1.4cm 1cm 1cm, clip=true]{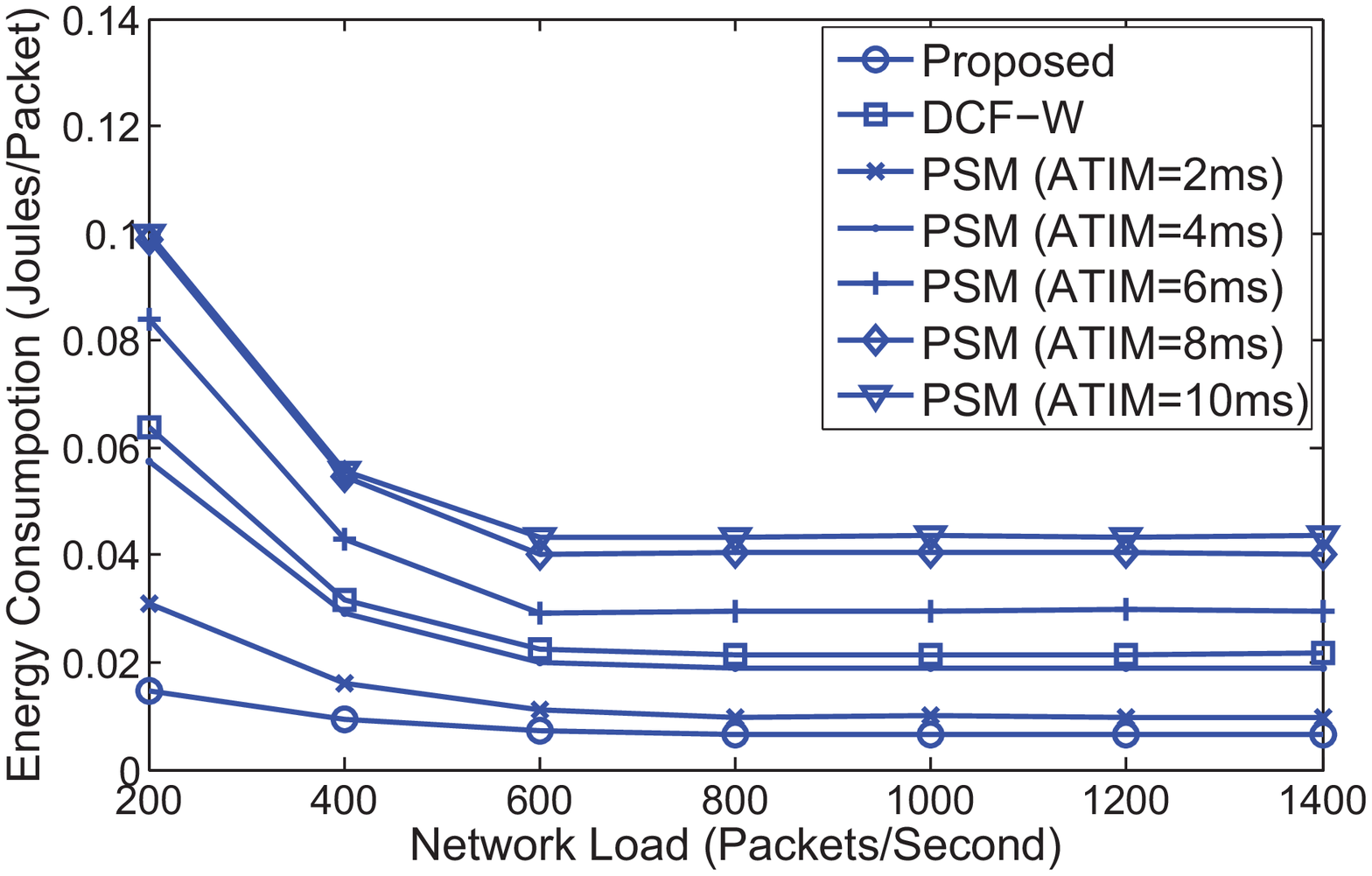}\label{E1}}
\subfigure[K=20]{\includegraphics[width=2.31in, trim=.4cm 1.4cm 1cm 1cm, clip=true]{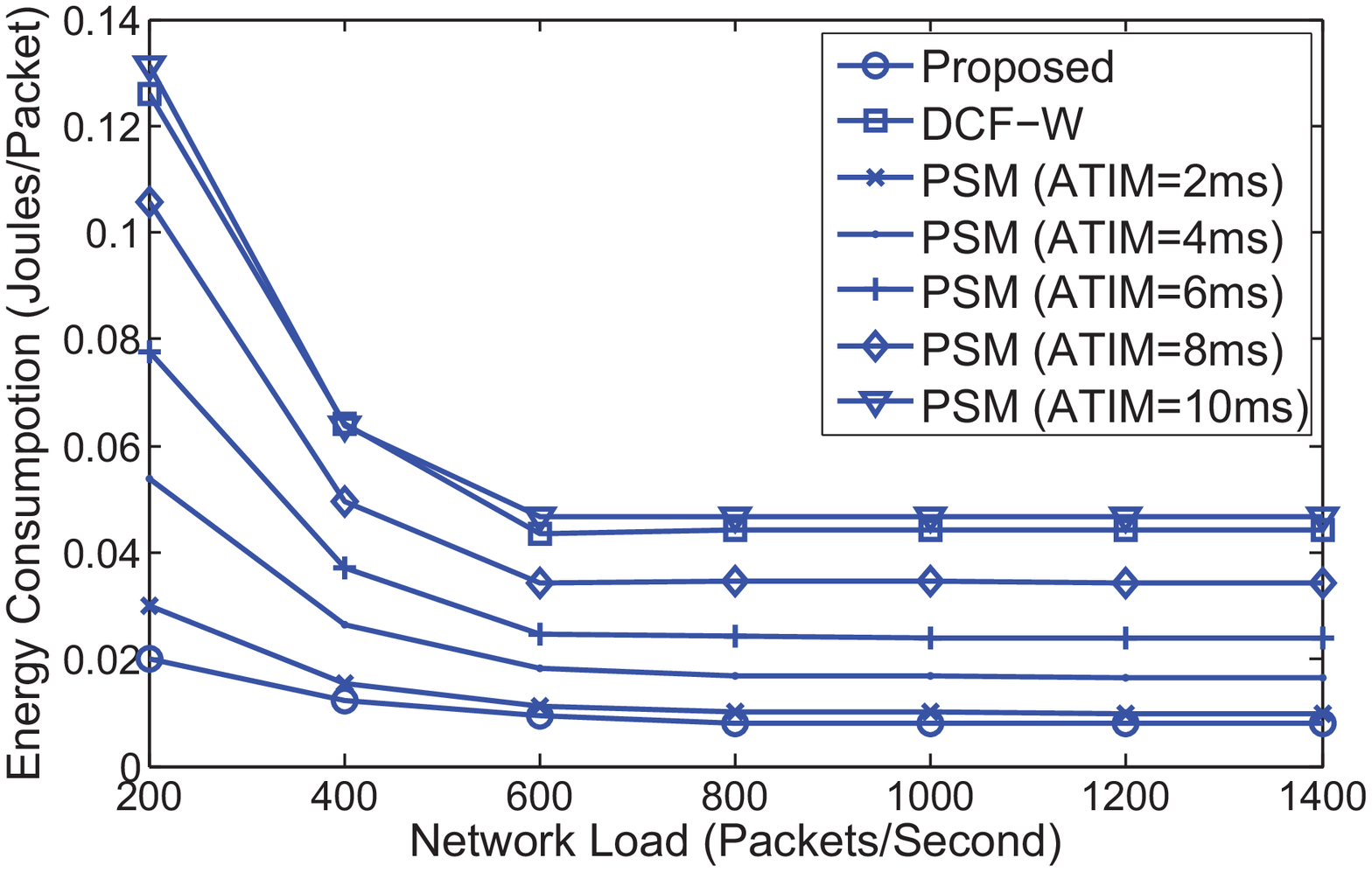} \label{E2}}
\subfigure[K=50]{\includegraphics[width=2.31in, trim=.4cm 1.4cm 1cm 1cm, clip=true]{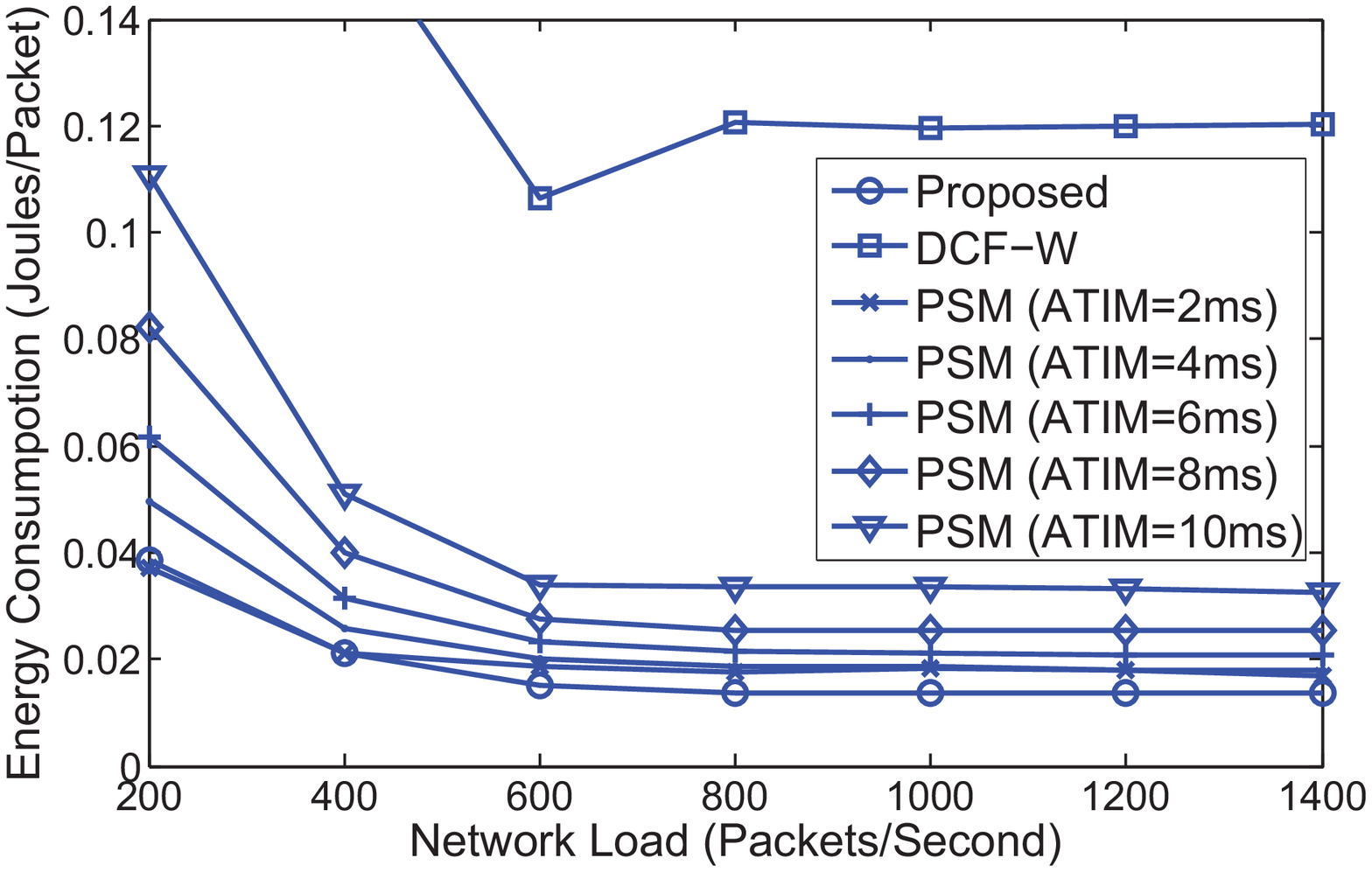}\label{E5} }
\caption{Energy consumption per packet of the proposed scheme, PSM, and DCF-W.} \label{E}
\end{figure*}

\begin{figure*}
\centering
\subfigure[K=10]{\includegraphics[width=2.31in, trim=.4cm 1.4cm 1cm 1cm, clip=true]{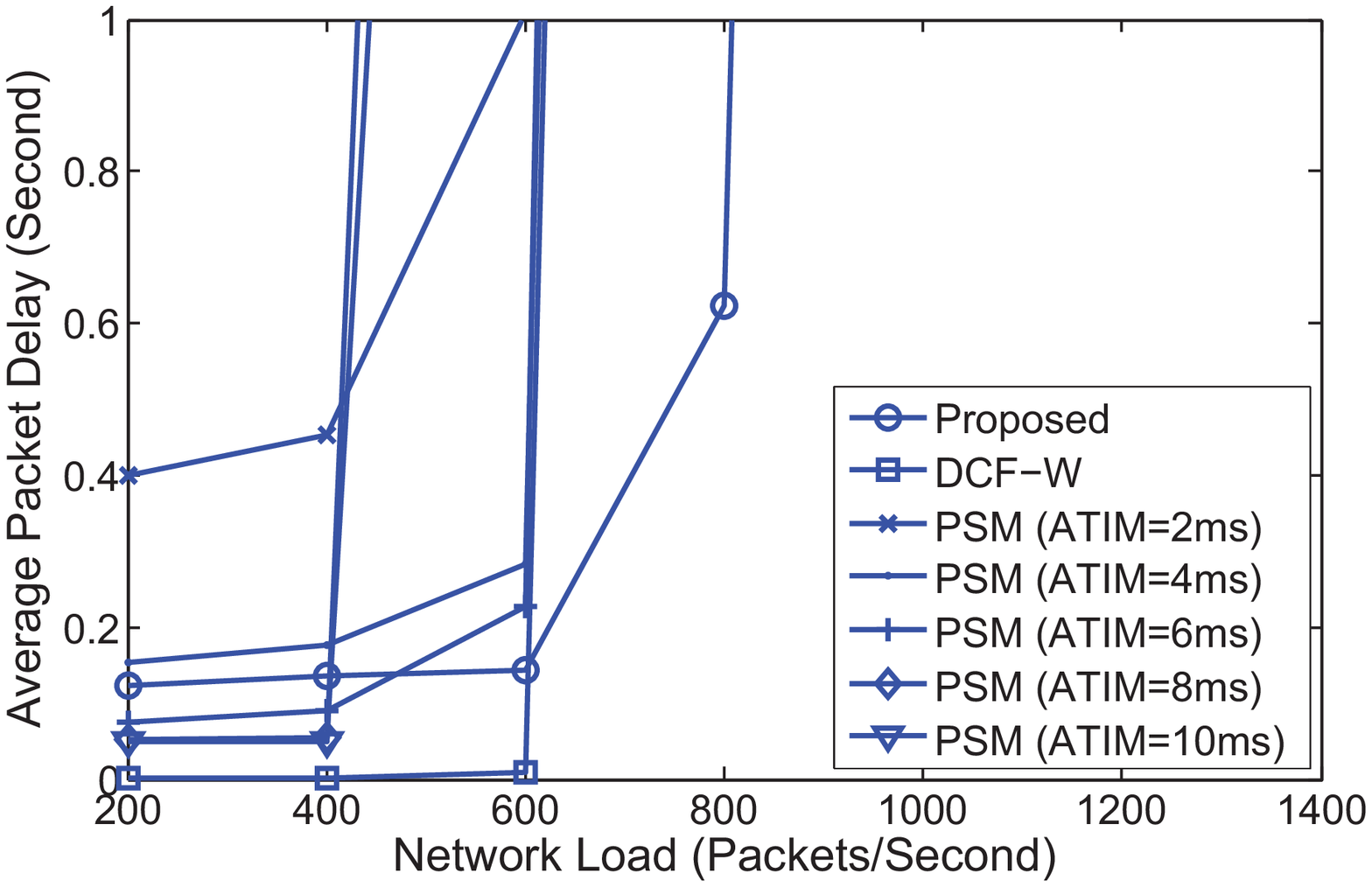}\label{D1}}
\subfigure[K=20]{\includegraphics[width=2.31in, trim=.4cm 1.4cm 1cm 1cm, clip=true]{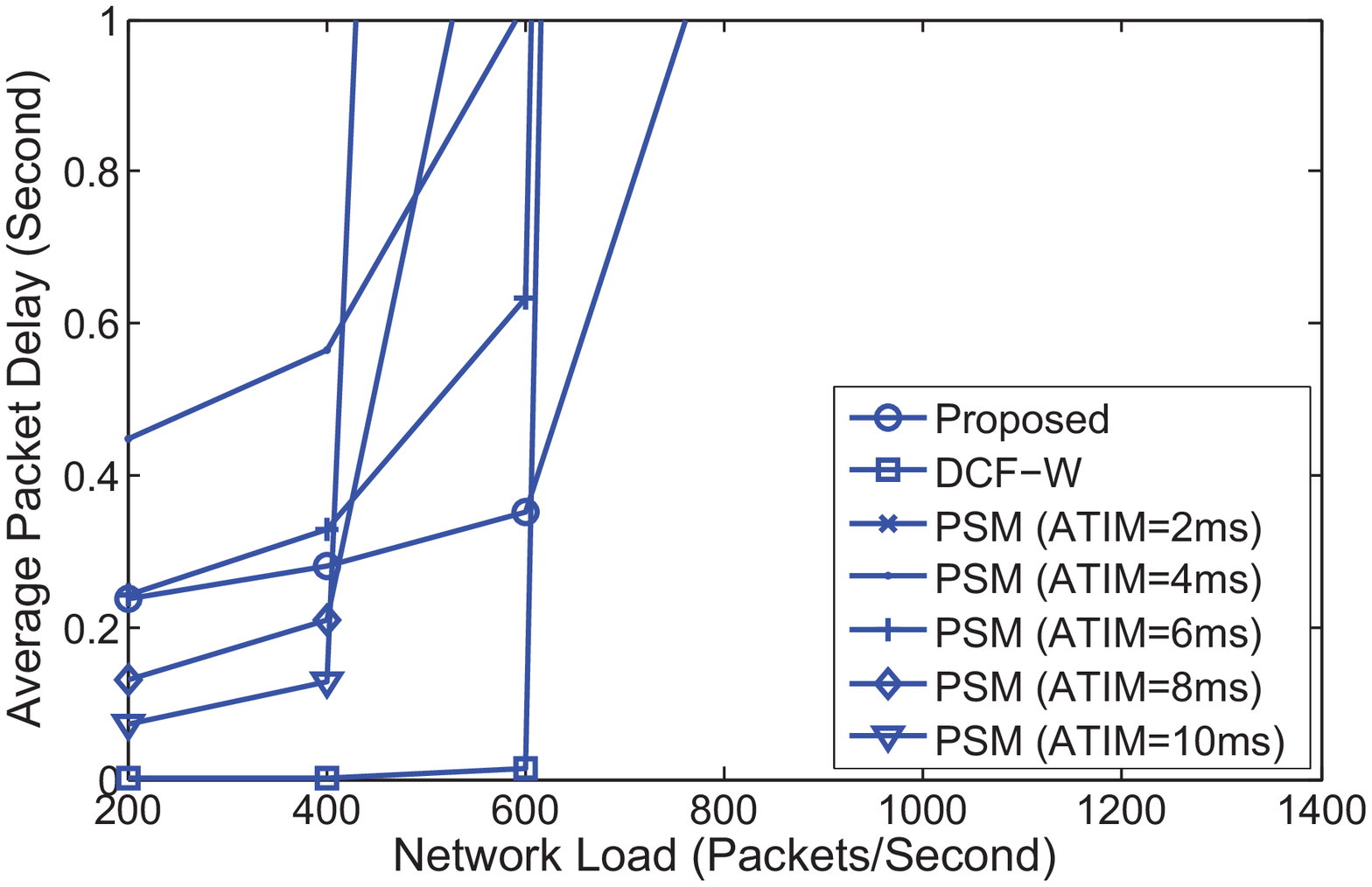} \label{D2}}
\subfigure[K=50]{\includegraphics[width=2.31in, trim=.4cm 1.4cm 1cm 1cm, clip=true]{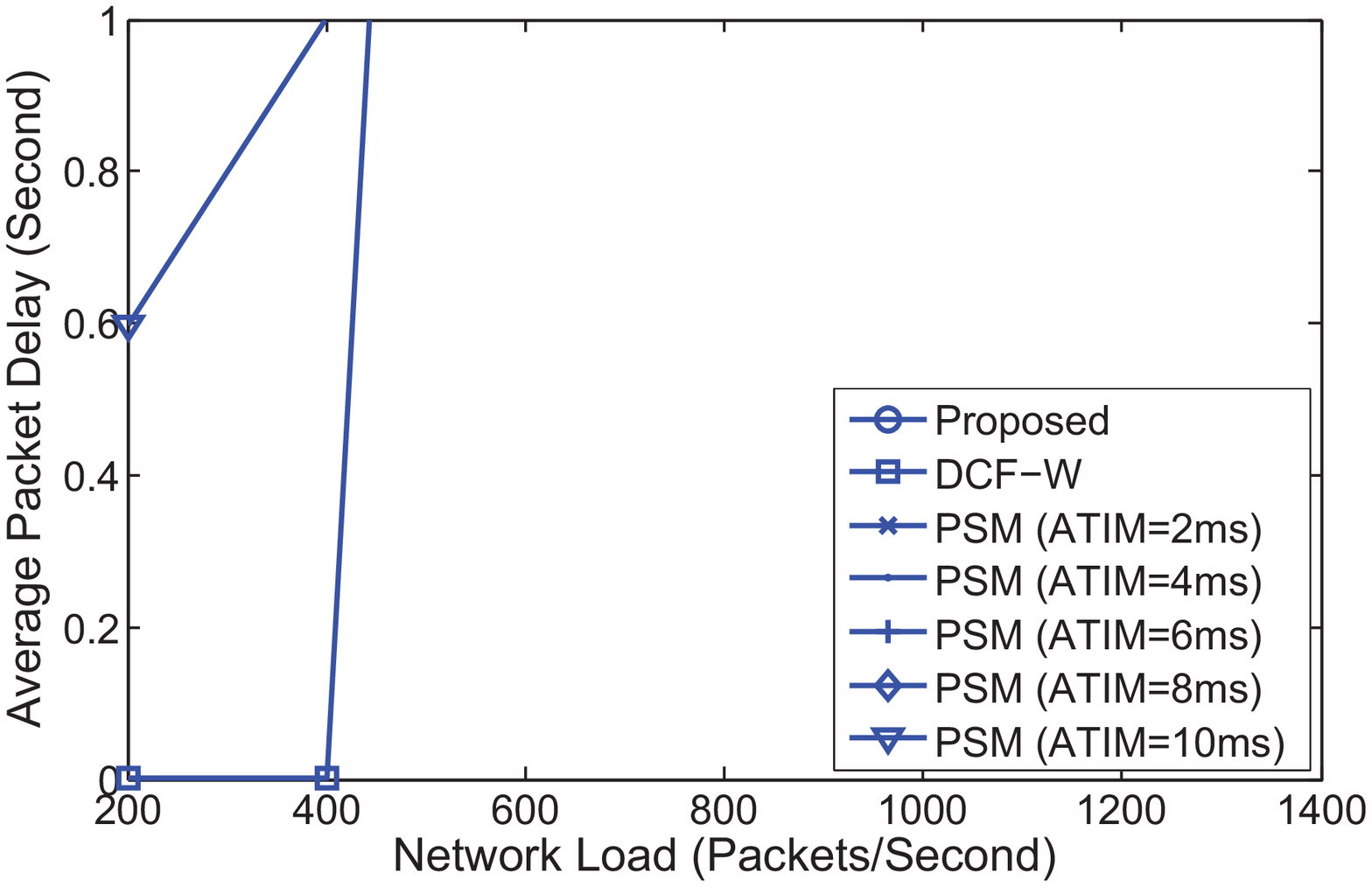} \label{D5}}
\caption{Average packet transmission delay of the proposed scheme, PSM, and DCF-W.} \label{D}
\end{figure*}

\subsection{Non-realtime traffic}
In this subsection, we consider only non-realtime traffic in the network and compare the throughput, energy consumption, and delay performance of our proposed scheme with the IEEE 802.11 DCF scheme without power saving (hereafter referred to as DCF-W) and in power saving mode (PSM).

The beacon interval $T_{nb}$ for both proposed scheme and PSM is set to $100ms$, which is the value specified for the PSM \cite{Standard}. Since the PSM performance significantly depends on the ATIM window size, we vary the ATIM window size from $2ms$ to $10ms$, which includes 4ms as specified in the standard \cite{Standard}. In the proposed scheme, the \begin{it}contention period\end{it} duration varies, depending on the \begin{it}contention-free period\end{it}. However, a minimum of $2ms$ is dedicated to the \begin{it}contention period\end{it} in each beacon interval to ensure that contending nodes can submit a request for the \begin{it}demand table\end{it} even when the network is overloaded. Other simulation parameters are given in Table \ref{Table}.

We compare the proposed scheme, DCF-W, and PSM as the network traffic load varies. Packets are generated at each node according to a Poisson process. The network load is defined as the aggregate packet generation rate in all the nodes. Three metrics are used as performance measures to compare the MAC schemes:
\begin{enumerate}
  \item \textit{Aggregate throughput,} which is defined as the total number of transmitted packets per second in the network;
  \item \textit{Energy consumption,} which is the average energy consumption per packet, and is calculated as the ratio of total energy consumption to the total number of transmitted packets in the network;
  \item \textit{Average packet delay,} which is the packet delay averaged over all the data packets transmitted in the network with packet delay being the duration from the instant that a packet is ready for transmission to the instant that the packet is successfully received at the receiver.
\end{enumerate}
Similar metrics are also used as performance measures in \cite{DPSM, On-demand, TMMAC, NAPman}, and \cite{E-MiLi}. Figures \ref{T}-\ref{D} show the aggregate throughput, energy consumption, and average packet delay of the proposed scheme, DCF-W and PSM versus the network load when there are $K=10,$ $20,$ $50$ nodes in the network. It is observed that the PSM performance depends on the ATIM window size. The PSM throughput is less sensitive to the ATIM window size when the network is light-loaded. However, as the traffic load increases, the ATIM window size significantly affects the PSM throughput. Generally, the ATIM window size should be adjusted based on the number of the contending nodes in the network. We consider a PSM scheme whose ATIM window size is dynamically adjusted to achieve the highest throughput (here after referred to as best-PSM), without imposing any overhead on the network. According to Figure \ref{T}, the ATIM size of best-PSM depends on the node number and is $2ms$, $4ms$ and $8ms$ for $K=10$, $K=20$ and $K=50$ nodes in the network respectively. In each scheme, the maximum achievable throughout decreases as the number of nodes increases, due to higher contention among nodes that causes more collision overhead. The results indicate that, for different network sizes ($K=10,K=20,$ and $K=50$ nodes), the proposed scheme provides 18\%-23\% higher throughput than the best-PSM and 27\%-43\% higher than the DCF-W.

\begin{figure}
\centering
\includegraphics[width=3.31in, trim=.5cm 1.3cm 1cm 1.4cm, clip=true]{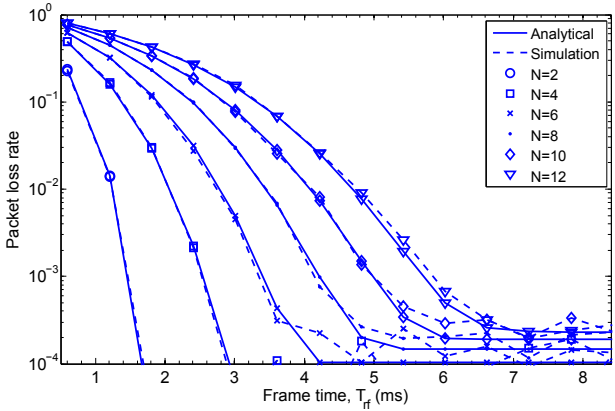}
\caption{Packet loss rate of realtime traffic.} \label{DF}
\end{figure}

\begin{figure}
\centering
\includegraphics[width=3.31in, trim=.5cm 1.3cm 1cm 1.4cm, clip=true]{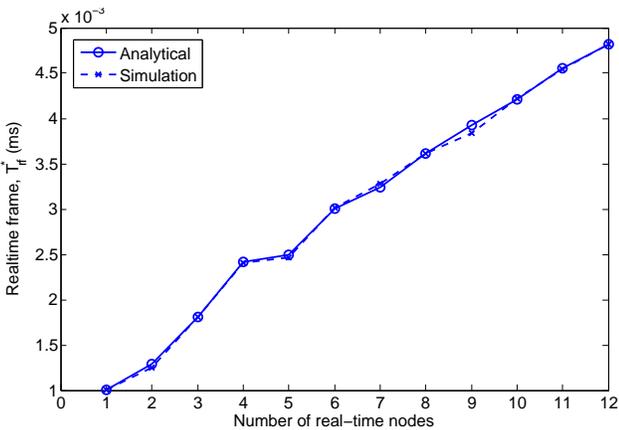}
\caption{Minimum required realtime frame duration to guarantee packet loss rate no larger than 1\%.} \label{RF}
\end{figure}

\begin{figure*}
\centering
\subfigure[Aggregate throughput of none-realtime nodes]{\includegraphics[width=2.31in, trim=.5cm 1.4cm 1cm 1cm, clip=true]{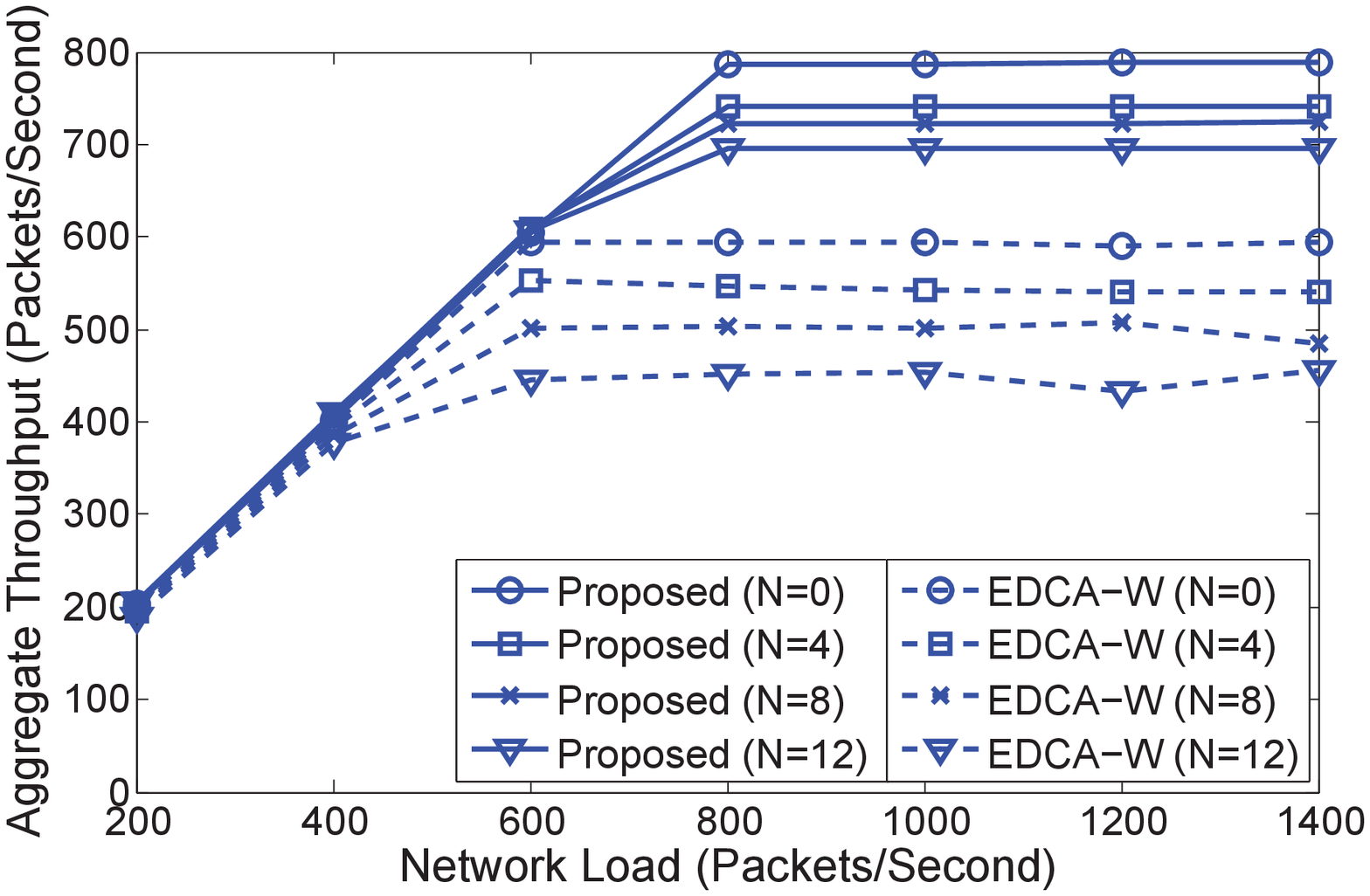}\label{TM}}
\subfigure[Packet loss rate at each real time node]{\includegraphics[width=2.31in, trim=.4cm 1.4cm 1cm 1cm, clip=true]{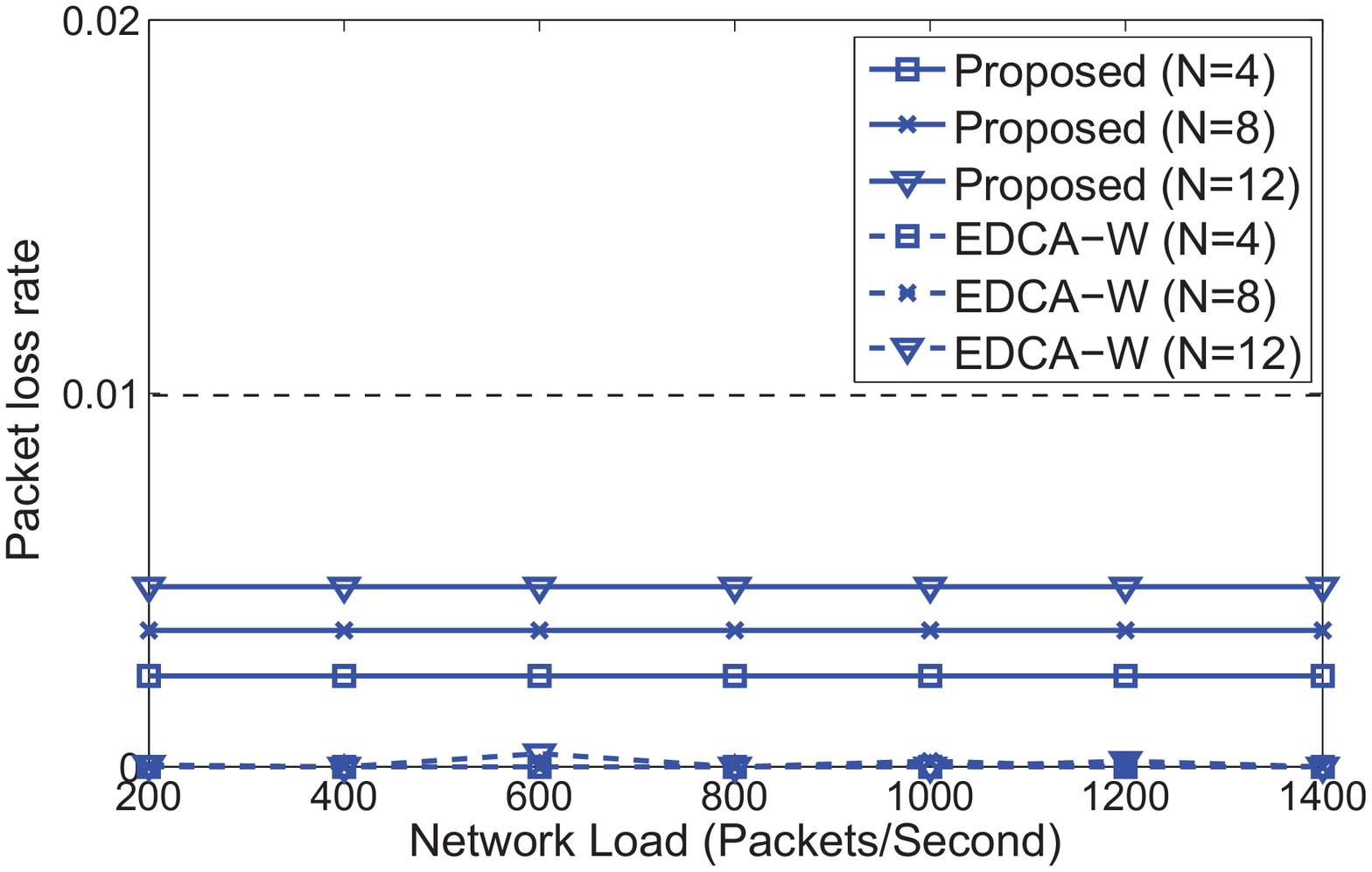}\label{PLRM} }
\subfigure[Total power consumption]{\includegraphics[width=2.31in, trim=.5cm 1.4cm 1cm 1cm, clip=true]{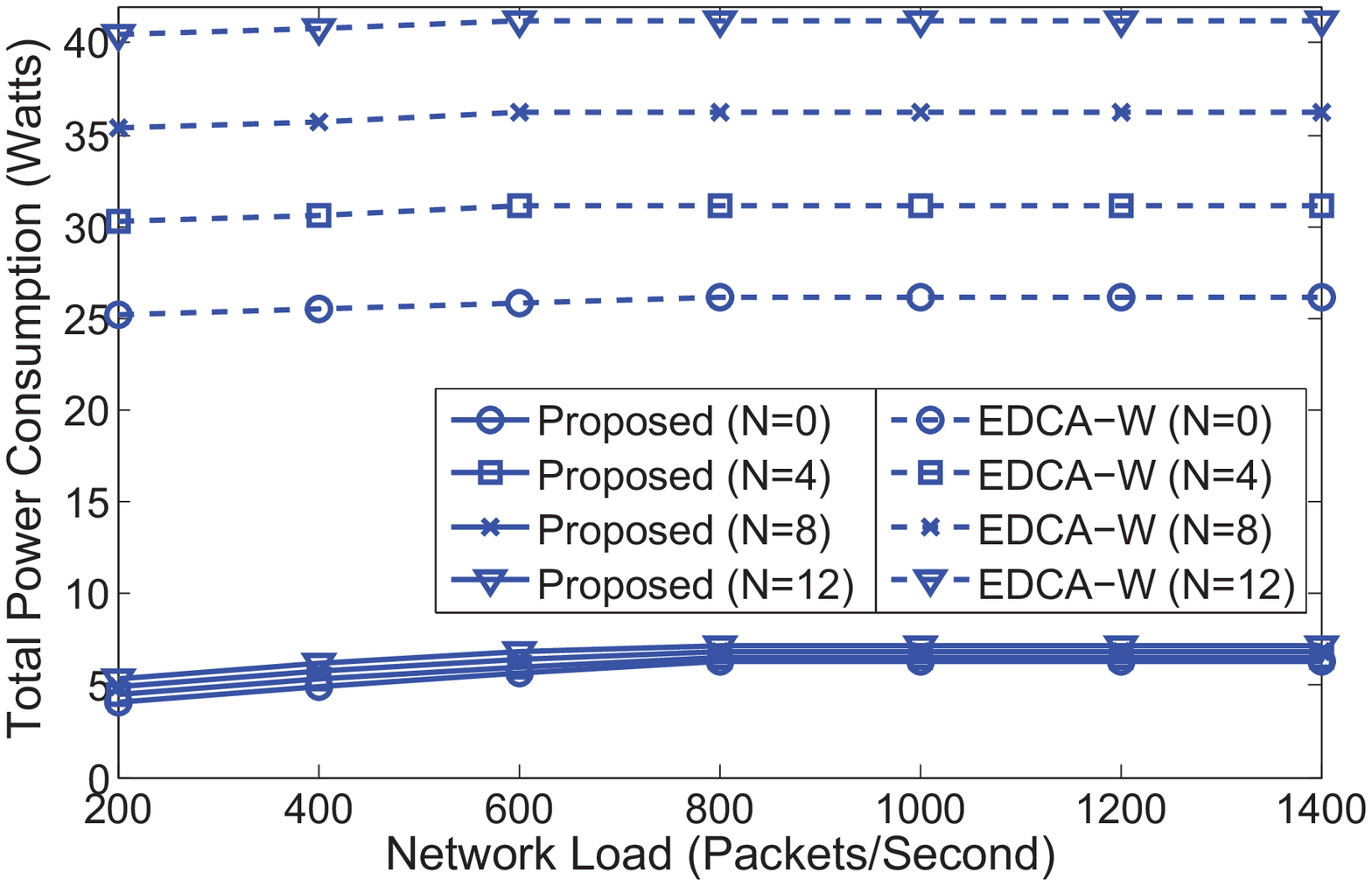} \label{PM}}
\caption{Performance of the proposed scheme, and EDCA-W for mixed realtime and non-realtime traffic (K=20 nodes).} \label{M}
\end{figure*}
Energy consumption per transmitted packet using different schemes is shown in Figure \ref{E}. As the number of nodes increases, the energy consumption per transmitted packet increases in each scheme due to more contention and collision among nodes. Although the total energy consumption in each scheme increases as the network load increases, all the schemes have the highest energy consumption per packet when the network load is the lowest. It is observed that the proposed scheme has a significantly lower energy consumption per transmitted packet, which is 48\%-55\% of the best-PSM.

Figure \ref{D} demonstrates that the proposed scheme and PSM have longer average packet delays than the DCF-W, as expected.  When the number of nodes increases and/or the network load increases, the average packet delay increases in each scheme. However, the proposed scheme provides a significantly lower average packet transmission delay as compared to the best-PSM.

\subsection{Realtime Traffic}
In this subsection, we calculate the minimum required frame duration that should be assigned to the realtime traffic to guarantee the required QoS, based on the analytical model presented in Section IV, and compare it with the simulation results. Consider the realtime traffic generated by voice codec G.711 (64 Kbps) at the nodes. Table \ref{Table2} lists the voice traffic parameters. Since a maximum end-to-end delay of $150 ms$ is recommended in \cite{ITU} for VoIP and video conferencing, we restrict the maximum tolerable delay of each voice packet to $50 ms$ in the wireless network. We set the beacon interval duration $T_{rb}=50ms$, and the maximum tolerable packet error rate of each voice call $\delta^*_{mac}=0.01$. Other parameters are given in Table \ref{Table}.

Figure \ref{DF} shows the packet loss rate versus the frame duration assigned to realtime traffic for different numbers of realtime nodes based on the analytical model and simulation. It is evident that there is a good match between the analytical and simulation results. The packet loss rate decreases almost exponentially as the frame duration increases. Figure \ref{RF} shows the minimum required frame time ($T^*_{rf}$) to guarantee the required packet loss rate, as the number of nodes with realtime traffic changes. The required channel time per realtime node (for constant average traffic load per node) decreases as the number of realtime nodes increases, due to a higher multiplexing gain.

\begin{table}[t]
    \begin{center}
    \caption{EDCA-W Parameters}
     \label{Table2}
        \begin{tabular}{l|c|c|c|c}

          \hline
          \hline
          Access category&CWmin&CWmax&AIFSN&MAX TXOP  \\
           \hline
           Realtime (Voice)&$3$&$7$&$2$&$1.504$ $ms$\\
           Non-realtime &$15$&$1023$&$3$&$0$\\
           (Best Effort)&&&&\\

          \hline
           \hline
        \end{tabular}
    \end{center}

\end{table}

\begin{figure*}[t]
\begin{equation}\label{8b}
  P_{X_1X_2X_3|S}(x_1,x_2,x_3|s)=\left\{
   \begin{array}{ll}
       1,&   n_2+n_3+n_4 \leq M, x_1=n_3, x_2=0, x_3=n_4;\\
       \frac{\dbinom{n_3}{x_1}\dbinom{n_4}{x_3}\dbinom{n_2}{M-x_1-x_3}}{\dbinom {n_2+n_3+n_4}{M}}, & n_2+n_3+n_4 > M, x_1+x_2=n_3,x_1+x_2+x_3=M,\\
        0, & \text{Otherwise}.
   \end{array}\right.
\end{equation}
\end{figure*}

\subsection{Mixed Realtime and Non-realtime Traffic}
Consider both realtime (voice) and non-realtime traffic in the network. We compare the performance of our proposed protocol with the Enhanced Distributed Channel Access without power saving mode (here after EDCA-W) which is defined in the IEEE 802.11e standard to provide the QoS guarantee for realtime traffic. We use default EDCA-W parameters as specified in the standard, given in Table \ref{Table2}. The network traffic load is evenly distributed between nodes with non-realtime traffic. Figure \ref{TM} shows the aggregate throughput of non-realtime traffic versus the network traffic load as the number of node with realtime traffic changes, with $K=20$. It is observed that the aggregate throughput of nodes with non-realtime traffic decreases as the number of nodes with realtime traffic increases in both proposed scheme and EDCA-W scheme. The packet loss rate of realtime traffic is illustrated in Figure \ref{PLRM}. Figure \ref{PM} shows the total power consumption of both realtime and non-realtime traffic. The results show that both proposed scheme and EDCA-W guarantee the required packet loss rate of realtime traffic; However, the proposed scheme has much lower power consumption and provides significantly higher throughput for non-realtime traffic.

\section{Conclusion}
In this paper, we present a novel distributed MAC protocol for fully connected wireless networks. A temporary coordinator node (\begin{it}head node\end{it}) regulates transmission dynamically based on the network traffic load condition to reduce energy consumption. In the proposed protocol, nodes contend once to transmit a batch of packets, after that they will be assigned contention free times for transmission. We also present an analytical model to evaluate the performance of proposed scheme that enables us to determine the minimum required channel time to realtime traffic. We compare the proposed scheme with the DCF scheme of IEEE 802.11 without power saving (DCF-W), the EDCA scheme of IEEE 802.11e without power saving (EDCA-W), and a dynamic version of IEEE 802.11 power saving mechanism, where the ATIM window size is adjusted dynamically based on the network traffic load conditions to provide best throughput (best-PSM). The performance measures include aggregate throughput and average packet delay of non-realtime traffic, packet loss rate of realtime traffic, and the total energy consumption in the network. Numerical results show that the proposed scheme guarantees the QoS requirement of realtime traffic, significantly reduces the energy consumption, and considerably enhances the network performance in terms of throughput and packet transmission delay in comparison with the existing protocols. In comparison with the best-PSM, the newly proposed scheme provides around 20\% higher throughput, 50\% less energy consumption, and reduces the packet transmission delay by half. In this work, we focus on a fully connected network, in which at each instant only one node can transmit over the radio channel. An interesting further research direction is to extend the MAC in a multi-hop ad hoc network, where non-interfering links can transmit simultaneously and increasing concurrent transmissions is an important goal \cite{Concurrent-MAC}.

\appendices

\section{Derivation of $P_{X_1X_2X_3|S}(x_1,x_2,x_3|s)$}
Since all nodes in states $2$, $3$ and $4$ are equally likely to be scheduled for transmission, $P_{X_1X_2X_3|S}(x_1,x_2,x_3,s)$ can be obtained as in (\ref{8b}).

\section{Derivation of $P_{X_4|X_1X_2X_3S}(x_4|x_1,x_2,x_3,s)$}
In a \begin{it}contention period\end{it}, each contending node chooses a random backoff window size $w$, uniformly distributed between $[0,W-1]$, waits for $w$ mini-slots of idle channel time, and then transmits a transmission request. Let random vector $\bar{w}(s)=(w_1,..., w_{n_1})$ denote the back-off times chosen by the contending nodes, where $w_i$ is the backoff window size chosen by contending node $i\in\{1,2,..,n_1\}$ at system state $s$ and $w_i\in[0,W-1]$. Denote the set of all possible outcomes of random vector $\bar{w}(s)$ by $\mathcal{W}(s)$. Since nodes choose their random back-off times independently and uniformly between $[0,W-1]$, different outcomes have equal probability and the size of $\mathcal{W}(s)$ is
\begin{equation}\label{7aaa}
  |\mathcal{W}(s)|=W^{n_1}.
\end{equation}
Consider random vector $U=(X_4,X_4',W_l)$, where $X_4$ and $X_4'$ are the numbers of successful and collided transmissions respectively during the \begin{it}contention period\end{it}, and $W_l$ is the backoff window chosen by node(s) which sends the last transmission request in the \begin{it}contention period\end{it}. Let $I_{cp}(u,s)$ denote the number of mini-slots that channel is idle during the \begin{it}contention period\end{it} in system state $s$ when event $U=u \triangleq (x_4,x'_4,w_l)$ occurs. We have
\begin{equation}\label{11x}
  I_{cp}(u,s)=T_{cp}(s)-(x_4+x'_4)t_q.
\end{equation}
Contending nodes do not initiate transmissions (even if their back-offs reach zero) if there is not enough time remained in the \begin{it}contention period\end{it} to complete at least one request. Thus, in the system state $s$ and event $u$, a contending node that has chosen backoff time $w > T_{cp}(s)-t_q-(x_4+x'_4)t_q=I_{cp}(u,s)-t_q$ does not start transmission. Also event $u$ is feasible in system state $s$ if
\begin{equation}\label{12a}
  (x_4+x'_4)-1\leq w_l\leq w_x(u,s)=\min(I_{cp}(u,s)-t_q,W-1).
\end{equation}
Let $\mathcal{Y}(u,s)$ denote a subset of $\mathcal{W}(s)$ that leads to event $u$ in system state $s$. In the event, there are $x_4$ successful requests and $x'_4$ collisions, and the backoff window of node(s) that had the last transmission is $w_l$; thus, $x_4+x'_4-1$ transmissions have backoff $w\in{[0,w_l)}$, one transmission has backoff time $w=w_l$, none of the other nodes has backoff $w\in{(w_l,w_x(u,s)]}$ and all other contending nodes have backoff $w\in{(w_x(u,s), W-1]}$. In addition, one node transmits at each successful transmission and at least two nodes in a collision. Therefore, the size of $\mathcal{Y}(u,s)$ is
\begin{multline}\label{12bbb}
  |\mathcal{Y}(u,s)|=\binom{w_l}{x_4+x'_4-1} \binom{x_4+x'_4}{x_4}\\ \frac{n_1!}{(n_1-x_4-2x'_4)!2^{x'_4}}  \Big(W-1-w_x(u,s)+x'_4\Big)^{n_1-x_4-2x'_4}.
\end{multline}
Using (\ref{7aaa}) and (\ref{12bbb}), the probability of event $u$ at system state $s$ can be calculated by
\begin{equation}\label{12b}
  P_{U|S}(u|s)=\frac{|\mathcal{Y}(u,s)|}{|\mathcal{W}(s)|}
\end{equation}
and $P_{X_4|S}(x_4|s)$ can be calculated using (\ref{12b}) as
\begin{equation}\label{14}
  P_{X_4|S}(x_4|s)=\sum_{x'_4,w_l} P_{U|S}(u|s).
\end{equation}
Since the right side of (\ref{14}) is independent of $x_1$, $x_2$, and $x_3$, we have
\begin{equation}\label{14v}
  P_{X_4|X_1X_2X_3S}(x_4|x_1,x_2,x_3,s)=P_{X_4|S}(x_4|s).
\end{equation}

\section{Derivation of $P_{X_5...X_8|X_1...X_4S}(x_5,...,x_8|x_1,...,x_4,s)$}
Let $p$ denote the probability that a realtime call switches from the \textit{off} mode to the \textit{on} mode in one realtime beacon interval, and $q$ the probability that a realtime call switches from the \textit{on} mode to the \textit{off} mode in one realtime beacon interval. With the exponentially distributed \textit{on} and \textit{off} periods, we have
\begin{equation}\label{2}
  p=1-e^{-\frac{T_{rb}}{t_{off}}},\text{  }q=1-e^{-\frac{T_{rb}}{t_{on}}}.
\end{equation}
Since the realtime calls are independent \textit{on} and \textit{off} periods, the pmf of transition number due to call status changes can be calculated as
\begin{multline}\label{14b}
  P_{X_5...X_8|X_1...X_4S}(x_5,...,x_8|x_1,...,x_4,s)=\\P_{X_5|X_1...X_4S}(x_5|x_1,...,x_4,s)
  P_{X_6|X_1...X_4S}(x_6|x_1,...,x_4,s)\\P_{X_7|X_1...X_4S}(x_7|x_1,...,x_4,s)P_{X_8|X_1...X_4S}(x_8|x_1,...,x_4,s)
\end{multline}
where all terms at the right side of (\ref{14b}) have binomial distribution, as given by
\begin{multline}\label{4}
  P_{X_5|X_1...X_4S}(x_5|x_1,...,x_4,s)=\\ \binom{n_2+x_4}{x_5}q^{x_5}(1-q)^{n_2+x_4-x_5},
\end{multline}
\begin{multline}\label{5}
  P_{X_6|X_1...X_4S}(x_6|x_1,...,x_4,s)=\\ \binom{n_4+x_2-x_3}{x_6}p^{x_6}(1-p)^{n_4+x_2-x_3-x_6},
\end{multline}
\begin{multline}\label{6}
  P_{X_7|X_1...X_4S}(x_7|x_1,...,x_4,s)=\\ \binom{n_1-x_4}{x_7}q^{x_7}(1-q)^{n_1-x_4-x_7},
\end{multline}
\begin{multline}\label{61}
   P_{X_8|X_1...X_4S}(x_8|x_1,...,x_4,s)=\\ \binom{n_5+x_1+x_3}{x_8}
  p^{x_8}(1-p)^{n_5+x_1+x_3-x_8}.\\
\end{multline}

\ifCLASSOPTIONcaptionsoff
\fi

\bibliographystyle{IEEEtran}
\bibliography{myref}

\begin{IEEEbiography}[{\includegraphics[width=1in,height=1.25in,clip,keepaspectratio]{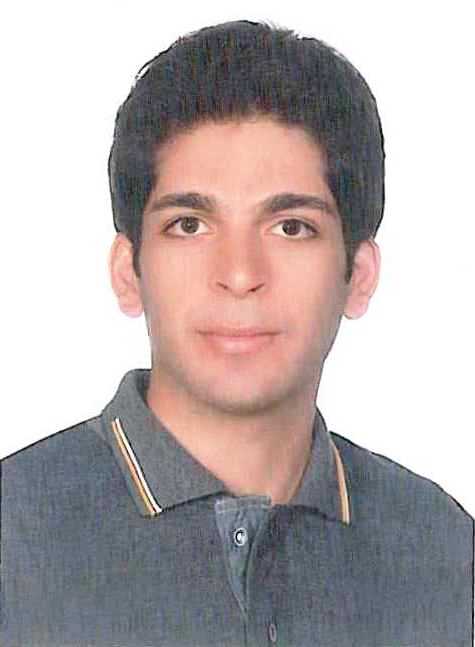}}]{Kamal Rahimi Malekshan} (S'11) received B.Sc. degree in electrical engineering from the University of Isfahan, Iran in 2008, and M.Sc. degree in electrical engineering from the University of Tehran, Iran in 2011. Since 2011, he has been working toward a Ph.D. degree at the Department of Electrical and Computer Engineering, University of Waterloo, ON, Canada.

His current research interest include medium access control (MAC), power management and transmission power control in wireless ad hoc networks.
\end{IEEEbiography}

\begin{IEEEbiography}[{\includegraphics[width=1in,height=1.25in,clip,keepaspectratio]{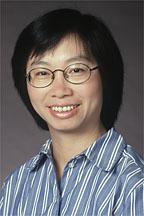}}]{Weihua Zhuang} (M'93-SM'01-F'08) has been with the Department of Electrical and Computer Engineering, University of Waterloo, Canada, since 1993, where she is a Professor and a Tier I Canada Research Chair in Wireless Communication Networks.
Her current research focuses on resource allocation and QoS provisioning in wireless networks. She is a co-recipient of the Best Paper Awards from the IEEE International Conference on Communications (ICC) 2007 and 2012, IEEE Multimedia Communications Technical Committee in 2011, IEEE Vehicular Technology Conference (VTC) Fall 2010, IEEE Wireless Communications and Networking Conference (WCNC) 2007 and 2010, and the International Conference on Heterogeneous Networking for Quality, Reliability, Security and Robustness (QShine) 2007 and 2008. She received the Outstanding
Performance Award 4 times since 2005 from the University of Waterloo, and the Premier’s Research Excellence Award in 2001 from the Ontario Government. Dr. Zhuang was the Editor-in-Chief of IEEE Transactions on Vehicular Technology (2007-2013), an editor of IEEE Transactions on Wireless Communications (2005-2009), the Technical Program Committee (TPC) Symposia Chair of the IEEE Globecom 2011, the TPC Co-Chair for Wireless Networks Symposium of the IEEE Globecom 2008, and an IEEE Communications Society Distinguished Lecturer (2008-2011). She is a Fellow of the IEEE, a Fellow of the Canadian Academy of Engineering (CAE), a Fellow of the Engineering Institute of Canada (EIC), and an elected member in the Board of Governors of the IEEE Vehicular Technology Society.
\end{IEEEbiography}

\begin{IEEEbiography}[{\includegraphics[width=1in,height=1.25in,clip,keepaspectratio]{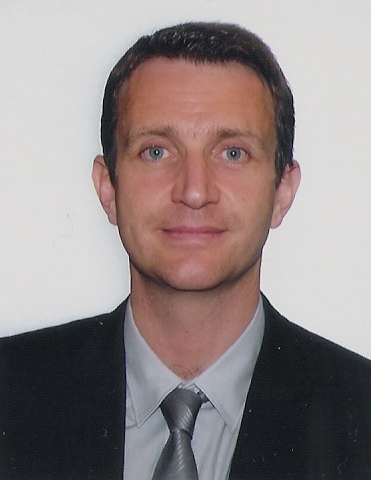}}]{Yves Lostanlen} (S'M98-M'01-SM'09) received the Dipl.-Ing (MSc EE) magna cum laude in 1996 from National Institute for Applied Sciences (INSA) in France. After three years of research at University College London and INSA he accomplished a European PhD summa cum laude in 2000. In 2009, he obtained a ScD (Habilitation) from University of Rennes, France and in 2013 he graduated from MIT Sloan School of Management, USA (Executive MBA.) Dr. Yves Lostanlen is currently CEO of SIRADEL North America and is based in Toronto, Canada, serving many top-tier companies in the ICT, Energy, Healthcare, Broadcast \& Media Industries: Government, policy makers, regulators, infrastructure operators and equipment manufacturers.
His current scientific interests are innovative technologies (hardware, software, data analytics) and services enabling energy-efficient infrastructure networks (telecom, energy, water) in under-developed regions in order to catalyze competitive advantage, productivity and growth.
Yves Lostanlen is also an Adjunct Professor in the Faculty of Applied Science and Engineering at the University of Toronto, Canada.
A senior member of IEEE, Yves Lostanlen has written more than 100 papers for international conferences, periodicals, book chapters and has been technical committee chairman, and session chairman at several international conferences. A frequent keynote speaker at international scientific and industrial conferences, Prof Lostanlen enjoys combining academic and industrial insights and technology and business constraints.
He received a "Young Scientist" Award" for two papers at the EuroEM 2000 conference.
\end{IEEEbiography}

\end{document}